\documentclass[prb,preprint]{revtex4}
\usepackage{graphicx}
\usepackage{bm}
\usepackage{amsmath}

\begin{document}
\title{Long-wave theory of bounded two-layer films with a free liquid-liquid interface:  Short- and long-time 
evolution}
\author{D. Merkt, A. Pototsky and M. Bestehorn}
\affiliation{Lehrstuhl f\"{u}r Theoretische Physik II\\ 
Brandenburgische Technische Universit\"at Cottbus\\ 
Erich-Weinert-Stra{\ss}e 1,\\
 D-03046 Cottbus, Germany\\
e-mail: merkt@physik.tu-cottbus.de}
\author{U. Thiele}
\affiliation{Max-Planck-Institut f\"ur Physik komplexer Systeme\\
N\"othnitzer Stra{\ss}e 38,\\
 D-01187 Dresden, Germany}
\begin{abstract}
We consider two-layers of immiscible liquids confined between an upper and
a lower rigid plate. The dynamics of the free liquid-liquid interface is
described for arbitrary amplitudes by a single evolution equation derived
from the basic hydrodynamic equations using long-wave approximation.
After giving the evolution equation in a general way, we focus
on interface instabilities driven by gravity, 
thermocapillary and electrostatic fields.
First, we study the linear stability discussing especially the conditions for
destabilizing the system by heating from above or below.
Second, we use a variational formulation of 
the evolution equation based on an energy functional to predict 
metastable states and the long-time pattern morphology (holes, drops or 
maze structures). Finally, fully nonlinear three-dimensional 
numerical integrations are performed to study the short- and long-time
evolution of the evolving patterns. Different coarsening modes are 
discussed and long-time scaling exponents are extracted.
\end{abstract}
\maketitle
%
%
%
\section{Introduction}
%
%
The study of stability properties and pattern formation in thin films is still an
important and not fully explored challenge in fluid dynamics. 
Reorganization processes of such films have remarkable applications in
chemical engineering, biological systems, or semiconductor industry. Besides
industrial aspects the computational advantage of studying thin films 
is obvious since one equation
for the surface is often sufficient to capture the basic dynamics. 
Due to the increase of computer power, pattern formation in systems far from
equilibrium can be investigated in more detail. This leads to consideration of
more and more complex systems which may show a rich variety of bifurcations
and allows for a more realistic modelling of fluid phenomena. 

Here, we consider thin two-layer films bounded by an upper and a lower rigid
plate. Using lubrication approximation 
a general long-wave interface evolution equation is derived 
that is valid for arbitrary amplitudes of interface deflections.
Pattern formation under the influence of gravity, thermocapillarity
and electrostatic forces is analysed. 

Lubrication or long-wave 
approximation is used for more than 100 years to describe the
evolution of thin films \cite{rey}.
In one-layer systems with a free interface
the dynamics of the surrounding gas is normally neglected and solely the liquid
determines the interface evolution. A simplified equation for the evolution of
the profile of the surface can be derived from the basic hydrodynamic
equations because the velocity is enslaved to the thickness profile. Several
mechanisms are known to destabilize an initially flat surface.
A survey of long-wave instabilities in one-layer systems is given by {\sc Oron
et al.} \cite{oron1}. A prominent example is the destabilization and
subsequent evolution of a non-flat surface profile due to
Marangoni flow caused by heating from below. It was first studied by 
{\sc Scriven} and {\sc Sterling} \cite{ScSt64} and classified by {\sc Goussis} and
{\sc Kelly} \cite{GoKe91} as the S-mode instability. The second mode, called P-mode, is a
short-wave instability without surface deflection and will not concern us
here. However, see {\sc Golovin et al.} \cite{GNP94} for a study of the interaction between short- and long-wave 
mode.
The linear and nonlinear behavior for an unstable thin liquid layer heated from below is
studied by {\sc Burelbach et al.} \cite{burel}. 
{\sc Deissler} and {\sc Oron} \cite{deiss} show the stabilizing effect of
thermocapillarity for a thin film at the underside of a cooled horizontal
plate which is gravitationally or Rayleigh-Taylor (RT) unstable.
The normally used linear dependence of surface tension on temperature (linear
Marangoni effect) is
replaced  by {\sc Oron} and {\sc Rosenau} \cite{oron2} by a quadratic one, thereby
inhibiting true film rupture. 
Three-dimensional simulations of the linear Marangoni effect are done by 
{\sc Oron} \cite{Oron00c} and for a wetting film by {\sc Bestehorn et al.} \cite{bes}. 
The former work concentrates on the evolution towards film rupture whereas the
latter system allows to explain the preference of drops or holes in
dependence of the film thickness. It also gives scaling exponents for the
long-time coarsening. In a recent work {\sc Thiele} and {\sc Knobloch}
\cite{thiele} show that the rich bifurcation structure for drop solutions
on a horizontal substrate is destroyed even by a rather small inclination of
the substrate. 

Mathematically related to thin heated films are ultrathin free surface 
films on horizontal substrates as first studied in long-wave approximation by
{\sc Ruckenstein} and {\sc Jain} \cite{RuJa73}.
These films may be unstable and dewet due to effective molecular
interactions that are incorporated into the governing equations 
in form of an additional pressure term.
This so-called disjoining pressure was introduced
by {\sc Derjaguin et al.} (for an overview see \cite{Isra92}). 
In the simplest case it results from the apolar London--van der Waals dispersion forces
\cite{RuJa73}. Open questions regarding dewetting in one-layer systems are 
summarized in Ref.~\cite{thiele2}.
Here, we will use a stabilizing van der Waals interaction to avoid
'true' film rupture for a heated film \cite{bes}.

The evolution of unstable thin films has a general
characteristics that is found as well for thin heated films as also
for ultrathin dewetting films.
One distinguishes between a short-time and a long-time
behavior. First, the flat film evolves into a large amplitude pattern whose
typical length scale can normally be determined by linear considerations. Often, this 
stage is called {\it initial film rupture} although the film may not rupture
totally, but an ultrathin film remains at the 'dry' parts.
The long-time behavior is characterized by an ongoing coarsening towards
patterns of longer and longer horizontal spatial scales.

Evidently, long-wave approximation is not only applicable for single liquid layers on a
solid substrate. The approach can be naturally extended towards systems
characterized by more than one layer. Taking, for example, two layers of
immiscible liquids on a solid horizontal substrate in a gas atmosphere yields
a pair of coupled
evolution equations for the liquid-liquid and the liquid-gas interfaces.
Such a system in presence of a surfactant and an evaporating upper liquid is
considered by {\sc Danov et al.} \cite{danov,danov2,paunov}.
Different pathways of dewetting induced by long-range van der Waals forces
are investigated by {\sc Pototsky et al.} \cite{poto}.
However, if such a two-layer system is bounded below {\it and}
above by rigid plates its behavior can be described by a single evolution
equation for the liquid-liquid interface. This kind of system is treated in the
present paper.

Although a general evolution equation was to our knowledge not yet given in the 
literature there exist a number of analyses for related systems. 
{\sc Yiantsios} and  {\sc Higgins} \cite{yian} 
investigate the Rayleigh-Taylor instability regarding an upper layer of
infinite thickness. They use lubrication approximation
for the lower liquid layer but not for the infinitely extended
upper one. They find that to leading order 
the dynamics of the upper layer can be neglected 
if the viscosities of both liquids are of the
same order of magnitude. In this way, they obtain an effective one-layer
interface evolution equation.

Marangoni effects in two superposed fluid layers are experimentally studied
by {\sc VanHook et al.} \cite{vanh}. They investigate long-wave as well as
short-wave thermocapillary instabilities. However, their
theoretical analysis neglects velocities in the upper layer and uses a
`two-layer Biot number' to take into account the thermal properties as well
as the thermal field in the upper layer. This
generalization of the Biot number used in \cite{oron1} also leads 
to an effective one-layer equation. 
A similar theory is used by {\sc Burgess et al.} \cite{burg} to explain the
stabilization of a Rayleigh-Taylor (RT) unstable oil-air system by heating
form below. 
Linear investigations of short- and long-wave Marangoni instabilities 
in two superposed liquid films are presented by {\sc Smith} \cite{smith}. 
Furthermore, {\sc Simanovskii} and {\sc Nepomnyashchy} \cite{sim,nep1} consider 
a two-layer system with thermocapillary effects.
They derive a weakly nonlinear interface equation in long-wave approximation
taking into account the dynamics in both
layers. Our linear results for the thermocapillary case 
can be directly compared to theirs. 
They show that the occurrence of
thermocapillary instabilities is not only determined by the direction of the
temperature gradient but also by the ratios of the viscosities 
and the layer thicknesses. In particular, they find that contrary to an 
one-layer system heating from above can act destabilizing.
Moreover, {\sc Tilley et. al} \cite{tilley} investigate two superposed
fluids in an inclined
channel with gravity and Marangoni effects. Their weakly nonlinear analysis 
reveals a modified Kuramoto-Sivashinsky equation with broken reflectional symmetry.

Two superposed dielectric fluids between two parallel plates 
are an appropriate system to investigate pattern formation through
electrohydrodynamic instabilities  since a vertically applied electric field
causes normal and tangential interface forces which depend strongly on the
dielectric fluid properties.   
{\sc Majumdar} and {\sc O'Neill} \cite{maju} propose an experimental method to
quantify surface tension via the measured critical voltage for the onset
of such an instability.  
{\sc Mohamed et al.} \cite{moha} 
investigate the destabilization of the interface using an Orr-Sommerfeld equation. 
Our linear results can be compared to theirs for quadratic velocity profiles
since long-wave approximation allows for quadratic velocity profiles only.  
A detailed analysis of different electrical fluid properties like the creation of free
charges at the interface, or the characterization of conducting and insulating
fluids is given by {\sc Melcher} and {\sc Smith} \cite{melch}.
Investigations of a free thin liquid jet with an axial applied electric field
are done by {\sc Savettaseranee et al.} \cite{savet}. They show that the electric
field stabilizes the film and can avoid rupture induced by attractive van der 
Waals forces.   
Experiments of {\sc Lin et al.} \cite{lin,lin2} using two layers of polymeric liquid
exposed to a vertical electric field suggest that linear considerations do
well capture the length scale found even in the long-time evolution.
Our nonlinear calculations confirm the validity of the linear theories.     

The present work focuses on two-layer films bounded by an upper and a lower rigid
plate as sketched in Fig.\,\ref{fig1}.  In Section~\ref{goveq}
we derive from the basic hydrodynamic
equations the interface evolution equation in lubrication or long-wave 
approximation for general layer properties. Keeping the normal and
tangential stresses at the interface in a general form, the resulting equation 
can be applied to the study of various body and interface forces. 
In the remaining paper we focus on gravity, thermocapillarity and 
electrostatic forces. This allows for the formulation of the problem in
variational form using a Lyapunov functional. 
Since the free energy density is a function of the interface $h$
only, the long-time evolution can be predicted, i.e.\ whether holes, drops or
maze structures are expected for $t\to\infty$. Further on we discuss 
the occurrence of metastable states. 
In Section~\ref{resdis}, we perform linear and nonlinear analyses of 
the derived equation.
First, we show that gravitation and thermocapillarity may act
stabilizing as well as destabilizing depending on material and system 
parameters.
Furthermore, we integrate the fully nonlinear evolution equation in three dimensions
numerically and study the short-time as well as the long-time
evolution.  For the latter, different coarsening modes and the long-time
scaling are discussed.
We summarize our results in Section~\ref{summ}  and point out possible applications,
in particular the influence of electrostatic fields on dewetting. 
In the Appendix we discuss shortly the subtle occurrence of an additional 
mean flow field if the system is extended from 2D to 3D. 
%
\section{Governing equations}
\label{goveq}
%
%
We consider a two-layer system bounded by a rigid upper and lower plate with
a system thickness $d$ and a flat film interface with height $h_0$
(Fig.\,\ref{fig1}). Instabilities lead to a time and space dependent interface
profile $h(x,y,t)$.
\subsection{Evolution equation}
The  required material parameters for incompressible fluids are the densities
$\rho_i$ and the dynamic viscosities $\mu_i$. The subscripts $i=1$ and $i=2$ denote
liquid $1$ (lower layer) and liquid $2$ (upper layer), respectively. 
In long-wave approximation the governing equations are found by a perturbation
series in powers of the small parameter $\epsilon$ \cite{oron1}. We write in
two dimensions (2D)
\begin{subequations}\label{exp}
\begin{eqnarray}
	u_i & = & u_{i_0} + \epsilon u_{i_1} + \epsilon^2 u_{i_2} + \dots\\
	w_i & = & w_{i_0} + \epsilon w_{i_1} + \epsilon^2 w_{i_2} + \dots\\
	P_i & = & P_{i_0} + \epsilon P_{i_1} + \epsilon^2 P_{i_2} + \dots
\end{eqnarray}
\end{subequations}
where $u_i$ and $w_i$ stand for the $x$- and  $z$-components of the
velocities, respectively. The small parameter 
$\epsilon = \frac{2\pi h_0}{\Lambda} \ll 1$ reflects the fact
that the interface deflections are long scale, i.e.\ the mean film thickness $h_0$ is
small compared to the typical lateral length scale $\Lambda$. 

Taking the large difference in length scales into account,
it is natural to scale the system lengths like $z= h_0 z'$ and $x =
\frac{h_0}{\epsilon} x'$, the velocities like $u_i=u_0 u_i'$ and $w_i =
\epsilon u_0 w_i'$, the time like $t=\frac{h_0}{u_0\epsilon} t'$,
the pressures like $P_i = \frac{\mu_1 u_0}{h_0\epsilon} P_i'$, the body forces like
$\Phi_i = \frac{\mu_1 u_0}{h_0\epsilon} \Phi_i'$, and the normal and tangential
interface forces like $\Pi = \frac{\mu_1 u_0}{h_0} \Pi'$ and $\tau =
\frac{\mu_1 u_0}{h_0} \tau'$. The primes denote the dimensionless
variables. $u_0$ is a reference velocity of fluid~1 parallel to the substrate. 

Starting from the two-dimensional incompressible Navier-Stokes equations
and the continuity equations for both liquid layers we derive the dimensionless equations
in zeroth order in $\epsilon$. After substituting Eqs.\,(\ref{exp}) in the 
governing equations and neglecting all terms of $O(\epsilon)$ or smaller, we
drop the primes and the subscript zero and obtain for the scaled quantities  
in the lower layer, $0\le z\le h(x,t)$
\begin{subequations}
\begin{eqnarray}
	\partial_z^2 u_1 & = & \partial_x \tilde{P_1}\label{e1}\\
	\partial_z \tilde{P_1} & = & 0\label{e2}\\
	\partial_x u_1 +\partial_z w_1 & = & 0\label{e3}
\end{eqnarray}
and in the upper layer, $h(x,t) \le z\le d$
\begin{eqnarray}
	\mu \partial_z^2 u_2 & = & \partial_x \tilde{P_2}\label{e4}\\
	\partial_z \tilde{P_2} & = & 0\label{e5}\\
	\partial_x u_2 +\partial_z w_2 & = & 0.\label{e6}
\end{eqnarray}
\end{subequations}
The variables 
\begin{equation}\label{ptilde}
	\tilde{P_1} = P_1 + \Phi_1 \qquad\mbox{and}\qquad \tilde{P_2} = P_2 + \Phi_2
\end{equation}
stand for reduced pressures which are the sum of the hydrostatic pressure
$P_i$ and the potential of the conservative body force $\Phi_i$ (e.g.\ gravity
force). The parameter $\mu=\mu_2/\mu_1$ represents the ratio of the dynamic
viscosities.
The boundary conditions at the lower and the upper boundaries read
\begin{subequations}\label{b1}
\begin{equation}
	u_{1} = 0,\quad\quad w_{1} =0\quad\quad\quad\quad\quad\quad
	 \mbox{at}\quad z=0\\
\end{equation}
and
\begin{equation}
	u_{2} = 0,\quad\quad w_{2} =0\quad\quad\quad\quad\quad\quad
	 \mbox{at}\quad z=d.\\
\end{equation}
\end{subequations}
The resulting interface conditions in zeroth order at $z=h(x,t)$ are
\begin{subequations}
\begin{eqnarray}
	\tilde{P_1} - \tilde{P_2} & = &  {\cal N} +\Phi \label{i1}\\
	\partial_z u_1 - \mu\partial_z u_2 & = &  {\cal T}\label{i2}\\
	\partial_t h + u_1\partial_x h & = & w_1\label{i3}\\
	\partial_t h + u_2\partial_x h & = & w_2 \label{i4}\\
	u_1 & = & u_2\label{i5}\\
	w_1 & = & w_2\label{i6}
\end{eqnarray}\end{subequations}
In the remainder of the paper we use the abbreviations ${\cal N}$  for the
normal forces Eq.\,(\ref{i1}), ${\cal T}$ for the tangential forces at the
interface Eq.\,(\ref{i2}) and $\Phi =\Phi_1 - \Phi_2$ for the body force
potential. We mention that the occurrence of the body force potential $\Phi$ in
Eq.\,(\ref{i1}) is caused by using reduced pressures Eq.\,(\ref{ptilde}). 

Because $\tilde{P_1}$ and $\tilde{P_2}$ do not depend on $z$ (Eqs.\,(\ref{e2}) and
(\ref{e5})) one can integrate Eqs.\,(\ref{e1}) and (\ref{e4}) in $z$. With the
boundary conditions Eq.\,(\ref{b1}) and the interface conditions Eqs.\,(\ref{i2}),
(\ref{i5}) the $x$-components of the velocities take the explicit form
\begin{subequations}\label{velu}
\begin{eqnarray}
	u_1(x,z,t) & = &\frac{1}{2}\partial_x\tilde{P_1}\; z^2 +
		\left(-h \partial_x {\cal N} - h \partial_x\Phi  + B + {\cal T} \right)z\label{u1}\\
	u_2(x,z,t) & = &\frac{1}{2\mu}(\partial_x\tilde{P_1} - \partial_x {\cal N} 
	-\partial_x\Phi)\; \left(z^2 - d^2\right) +
	\left(\frac{1}{\mu}B\right)(z-d)
	\\[1em]
	\mbox{with\hspace{1em}}
	B & = & \frac{1}{2}\frac{-(\partial_x\tilde{P_1} - 2\partial_x {\cal N} 
	- 2\partial_x\Phi)\mu h^2 + (\partial_x\tilde{P_1} - \partial_x {\cal N} 
	-\partial_x\Phi)(h^2-d^2) -2\mu {\cal T} h}{(\mu -1) h +d}\nonumber
\end{eqnarray}
\end{subequations}
where we used Eq.\,(\ref{i1}) to express $\partial_x \tilde{P_2}$ as a function
of $\partial_x\tilde{P_1}$. 

Next, we derive the evolution equation for the interface profile $h(x,t)$ and
an explicit formula for the pressure $\partial_x\tilde{P_1}$. To do so the
continuity equation Eq.\,(\ref{e3}) is integrated in $z$-direction. With the
interface condition Eq.\,(\ref{i3}) and the chain rule we find
\begin{equation}\label{ev1}
	\partial_t h + \partial_x \int_0^{h(x,t)}{u_1(x,z,t)\,dz} = 0.
\end{equation}
A similar evolution equation for $h(x,t)$ is also derived for the upper layer
using Eqs.\,(\ref{e6}) and (\ref{i4}). Then Eq.\,(\ref{i6}) yields the identity 
\begin{equation}\label{int}
	\partial_x\left(\int_0^{h(x,t)}{u_1(x,z,t)\,dz} +
\int_{h(x,t)}^d{u_2(x,z,t)\,dz} \right) = 0. 
\end{equation}
To obtain $\tilde{P_1}$, Eq.\,(\ref{int}) is integrated in $x$
setting the integration constant to zero without loss of generality. This can
be done if there is no additional lateral driving force as, for instance, in
an inclined system. The resulting equation is solved  using Eq.\,(\ref{b1}).
The resulting pressure gradient is
\begin{eqnarray}\label{p1}
	\partial_x\tilde{P_1} & = & F_1(h)\partial_x\left( {\cal N} + \Phi\right) + F_2(h){\cal T} 
\end{eqnarray}
with
\begin{subequations}
\begin{eqnarray}
	F_1(h) & = & \frac{1}{D}(d-h)^2(h\mu(4d-h) + (d-h)^2)\label{f1}\\
	F_2(h) & = & \frac{6\mu d h }{D}(d -h)\\
	D & = &  (d-h)^4 + h\mu(h^3(\mu-2)+4dh^2-6d^2h + 4d^3).  \label{cf}	
\end{eqnarray}
\end{subequations}
The evolution equation Eq.\,(\ref{ev1}) can be written (with Eq.\,(\ref{u1})) as
\begin{equation}\label{ev}	
	\partial_t h   = 
	\partial_x\left[ Q_1(h)\,\partial_x \left({\cal N} +  \Phi \right)
	+Q_2(h)\,{\cal T}\right].	
\end{equation}
Using the same procedure we derive a similar equation for three dimensions
\begin{equation}\label{ev3d}	
	\partial_t h   = 
	\nabla\cdot\left[ Q_1(h)\,\nabla( {\cal N} + \Phi)
	+Q_2(h)\,\vec{\cal T}\right]
\end{equation}
where $\nabla=(\partial_x,\partial_y)$.
We note that in 3D an additional field occurs which can not be 
expressed as a function of $h$ analytically. 
The importance and
the equation of this mean flow field are discussed
in the appendix.
However, we neglect the mean flow in the following.

The mobilities are
\begin{subequations}\label{m}
\begin{eqnarray}
	Q_1(h) & = & \frac{h^3\,(d-h)^3}{3D}\,[d+h(\mu-1)]\\
	Q_2(h) & = & \frac{h^2\,(d-h)^2}{2D}\,[h^2(\mu-1)-d(d-2h)].\label{mq2}
\end{eqnarray}
\end{subequations}
The mobility $Q_1(h)$ is positive for all values $\mu>0$ and  $d>0$. It
vanishes for $h=0$ and $h=d$. However, $Q_2(h)$ always changes its sign at
\begin{equation}\label{zc}
h_c=\frac{d}{\sqrt{\mu}+1}.
\end{equation}
The mobility
$Q_2(h)$ does only exist in a system with shear-stress. In Fig.\,\ref{fig2}
the mobilities are plotted for the parameters of Tab.\,\ref{tabmat} with
$d=1.3$.
There the change of sign of $Q_2(h)$ occurs at $h_c\approx 0.91$.

In an one-layer system the viscosity of the upper gas layer is neglected. 
Therefore, taking the limit $\mu\to 0$ of the mobilities Eqs.\,(\ref{m}) give
the correct one-layer mobilities 
\begin{equation}\label{mlim}
	Q_1(h)_{\lim} = \frac{1}{3}h^3, \hspace*{3em}\mbox{and}\hspace*{3em} Q_2(h)_{\lim} = -\frac{1}{2}h^2,
\end{equation}
and the evolution equation Eq.\,(\ref{ev3d}) takes the well known form of the
thin film equation for a single layer \cite{oron1}
\begin{equation}
	\partial_t h = -\nabla\left[-\frac{1}{3}h^3\nabla{\tilde P}_{1\,\lim} + \frac{1}{2}h^2{\cal T}_{\lim} \right].
\end{equation}
The limit of the pressure is
\begin{equation}
	{\tilde P}_{1\,\lim}= P_1 + \Phi_1
\end{equation}
since $\nabla{\tilde P}_2= 0$ (see Eq.\,(\ref{e4})) and  $\Phi_2$ can be
neglected as can be seen, for example,  for gravity forces where
$\Phi_i\propto \rho_i$  and $\rho_2\ll\rho_1$  
leads directly to $\Phi_1\gg\Phi_2$. The same limit is reached by 
increasing the system thickness $d\to\infty$.
\subsection{Specific effects}\label{effect}
\subsubsection{Gravitation and surface tension}
Incorporation of gravitation and surface tension (capillarity) provides
the body force potential and  the normal interface force
\begin{subequations}\label{ntp}
\begin{eqnarray}
	\Phi_G & = & (1-\rho) G h\\
	\mbox{and\hspace{1em}}
	{\cal N} & = & -C^{-1}\nabla^2 h\label{ntp2},
\end{eqnarray}
\end{subequations}
respectively.
Thereby, $G=(\rho_1 g h_0^2)/(\mu_1 u_0)$ is the gravity number,
$\rho=\rho_2/\rho_1$ denotes the ratio of densities, $C^{-1}=\epsilon^3
\sigma/(\mu_1 u_0)$ denotes the capillary number and $\sigma$ is the
dimensional liquid-liquid surface tension.
\subsubsection{Thermocapillarity}
The equations for the non-dimensional temperatures $\Theta_i$ (energy
equations) read in zeroth order in $\epsilon$
\begin{subequations}\label{t1}
\begin{eqnarray}
	\partial_z^2 \Theta_1 & = & 0\\
	\partial_z^2 \Theta_2 & = & 0.
\end{eqnarray}
\end{subequations}
The non-dimensional temperature is defined by
\begin{equation}
	\Theta_i = \frac{T_i-T_{u}}{T_{l}-T_{u}}
\end{equation}
where $T_{u}$ and $T_l$ refer to the temperature at the upper and lower plate,
respectively.
The boundary conditions at the lower and upper rigid plate are
\begin{subequations}\label{bt1}
\begin{eqnarray}
	\Theta_1  &  = &  1 \quad\quad\quad\quad\quad\quad \mbox{at}\quad\quad z=0 \\
	\mbox{and\hspace{1em}}
	\Theta_2  &  =  & 0 \quad\quad\quad\quad\quad\quad \mbox{at}\quad\quad z=d
\end{eqnarray}
\end{subequations}
and the interface conditions at $z=h(x,t)$ read 
\begin{subequations}\label{it1}
\begin{eqnarray}
	\Theta_1 &   =  & \Theta_2\\
	\partial_z\Theta_1 & = & \lambda\partial_z\Theta_2,
\end{eqnarray}
\end{subequations}
where $\lambda =\lambda_2/\lambda_1$ is the ratio of the thermal conductivities.
Eqs.\,(\ref{t1}) together with Eqs.\,(\ref{bt1}) and (\ref{it1}) give the temperature fields
\begin{subequations}\label{tz1}
\begin{eqnarray}
	\Theta_1(z) & = & \frac{\lambda(h- z) + (d-h)}{(\lambda-1)h+d}\\
	\mbox{and}\qquad
	\Theta_2(z) & = & \frac{d-z}{(\lambda-1)h+d}.
\end{eqnarray}
\end{subequations}
Assuming an arbitrary dependence of the surface tension $\sigma$ on temperature, the
tangential interface condition Eq.\,(\ref{i2}) has the form ${\cal
T}=\nabla\Sigma$ where $\Sigma = \epsilon\sigma/(\mu_1 u_0)$ is the dimensionless
surface tension. Evaluation of $\nabla\Sigma$  at the
position $z=h(x,y,t)$  gives $\nabla\Sigma = \partial_\Theta
\Sigma\cdot\partial_h\Theta(h)\cdot\nabla h$. If the surface tension depends
linearly on temperature one gets 
\begin{equation}\label{stressb}
	{\cal T} = M \frac{\lambda d}{\left[(\lambda-1)h +d\right]^2}\nabla h
\end{equation}
where $M=(-\partial_T\sigma\,\Delta T \,\epsilon)/(\mu_1 u_0)$ is the
Marangoni number, $\partial_T\sigma$ is the change of surface tension with
temperature and $\Delta T = T_{l} - T_{u}$ is the applied temperature
difference. If the system is heated from below, $M$ is positive for most fluids
(normal thermocapillary effect).

To derive the one-layer limit of the thermocapillary force ${\cal T}_{\lim}$
one replaces $\lambda$  with a (conductive) Biot number $\lambda = Bi(d-1)$ in
Eq.\,(\ref{stressb}) and takes the limit $d\to \infty$. Considering the gas layer
as a semi-infinite layer, we get the usual one-layer expression \cite{oron1}
\begin{equation}
	{\cal T}_ {\lim} = M \frac{Bi}{\left(Bi\,h +1\right)^2}\nabla h,
\end{equation}
where $Bi\ll 1$.
We note that the `two-layer Biot number', $Bi_2$, introduced by {\sc VanHook
et al.} \cite{vanh} can be obtained by replacing
$\lambda =\frac{1-Bi_2 (d-1)}{1+Bi_2}$ in Eq.\,(\ref{stressb}) and taking the
limit $d\to 1$. This limit is valid since all dependencies on $d$ are
considered to be in $Bi_2$ and therefore the neglect of the upper layer
leads to $d \to 1$.
\subsubsection{Disjoining pressure}
To avoid film rupture at the two bounding plates we incorporate disjoining
pressures in the body force potential to model repelling stabilizing 
van der Waals forces \cite{will,bes}
\begin{subequations}
\begin{eqnarray}
	\Phi_{D_1} & = & \left.-\frac{H_1}{z^3}\right|_{z=h}\\
	\Phi_{D_2} & = & \left.-\frac{H_2}{(d-z)^3}\right|_{z=h}.
\end{eqnarray}
\end{subequations}
The parameters $H_1$ and $H_2$ are Hamaker constants representing 
the interaction of the lower plate with liquid 2 through liquid 1 and
of the upper plate with liquid 1 through liquid 2, respectively \cite{Isra92}. 
Thereby, we neglect a part of the forces between the lower (upper) fluid and the upper
(lower) substrate resulting from the finite thickness of the respective
layer. 
The Hamaker constants determine the macroscopic contact angles. Since large 
macroscopic contact angles violate the used long-wave approximation 
we use Hamaker constants which provide a small contact angle.
These corresponds to fluids which partially wet the substrates.  
Note, that for a very small distance between the
plates, i.e.\ if both layers are ultrathin with thicknesses below 100\,nm
the disjoining pressure has the most important influence and all
forces have to be included. Such systems are not the scope of the present
work, but see \cite{poto} for a related system.
\subsubsection{Electrostatic field}
An electric field applied in $z$-direction is another way to cause structure formation. 
Consider two dielectric fluids with permittivities $\varepsilon_1$ and
$\varepsilon_2$, respectively. The upper and lower plates serve as electrodes
and a voltage $U$ is applied.
The vertical components of the electric fields in fluid 1 and 2 read then in
zeroth order lubrication approximation
\begin{subequations}\label{efields}
\begin{eqnarray}
	E_1 & = & \frac{\varepsilon_2 U}{\varepsilon_2 h + \varepsilon_1(d-h)}
\\
	\mbox{and\hspace{1em}} E_2 & = & \frac{\varepsilon_1 U}{\varepsilon_2 h + \varepsilon_1(d-h)}\quad.
\end{eqnarray}
\end{subequations}
Horizontal components can be neglected at this order.
Using the electrohydrodynamic stress tensor \cite{lan} for  $\rho_i=$const., provides the
effective electrostatic pressure  at the interface by projecting  the stress
tensor two times on the normal vector
\begin{equation}
	p_{el} = \frac{1}{2}\varepsilon_0(\varepsilon_2-\varepsilon_1)E_1 E_2.
\end{equation}
Scaling the voltage like $U = U'\sqrt{\mu_1 u_0
h_0/\varepsilon_0\varepsilon_1\epsilon}$ and dropping the primes, leads in
zeroth order lubrication approximation to 
\begin{equation}
	{\cal N}_{el} =  \frac{1}{2}\frac{\varepsilon(\varepsilon -1) U^2}{(\varepsilon h + (d-h))^2},\end{equation}
where $\varepsilon_0$ denotes the permittivity of vacuum and $\varepsilon =
\varepsilon_2/\varepsilon_1$ is the ratio of permittivities. In zeroth order in
$\epsilon$ shear stresses ${\cal T}_{el}$ are not present. 
\subsection{Energy}
As for one-layer systems \cite{oron2} also here 
it is possible to express the r.h.s.\ of the evolution equation
Eq.\,(\ref{ev3d}) in variational form
\begin{equation}\label{ly1}
	\partial_t h = \nabla\cdot\left[Q_1(h)\;\nabla\frac{\delta E}{\delta h} \right].
\end{equation}
corresponding to the evolution equation of a conserved order parameter field
in a relaxational situation.
Incorporating the above mentioned effects the energy $E$ that corresponds to 
a Lyapunov functional can be written as
\begin{equation}\label{ly3}
	E = \int\!\int{dx\,dy\;\left[\frac{1}{2}C^{-1}(\nabla h)^2 + V(h)\right]}, 
\end{equation}
with
\begin{equation}\label{ly2}
	V(h) = \frac{1}{2}(1-\rho)G\,h^2 + \frac{H_1}{2} h^{-2} + \frac{H_2}{2}(d-h)^{-2}  + E_{th}(h) + E_{el}(h), 
\end{equation}
and 
\begin{eqnarray}
	E_{th}(h) & = & \frac{3 M}{2d\lambda(\mu-\lambda)^2}{\Bigg[}
	- \lambda^2(\mu-\lambda)^2 h\;\ln{(h)} + (\mu-\lambda)^2(d-h)\;\ln{(d-h)}\nonumber
	\\
	& &  + \lambda^2(\mu-1)(h(\mu-1)+d)\;\ln{(h(\mu-1)+d)}\nonumber
	\\
	& & + \left(\lambda^4 h - 2\lambda^3\mu h + \lambda^2\mu(2 h -d) + 2\lambda\mu(d-h) - \mu^2(d-h)\right)\;\ln{(h(\lambda-1)+d)}\Bigg]\nonumber
	\\
	\label{Menergy}
\end{eqnarray}
for thermocapillarity and
\begin{eqnarray}
	E_{el}(h) & = & -\frac{1}{2}\frac{\varepsilon U^2}{(\varepsilon-1)h+d}
\end{eqnarray}
for electrostatic fields.

It can easily be shown \cite{pen,oron1} that the Lyapunov functional $E$ is monotonously
decreasing in time ($\frac{d}{dt}E\le 0$), if the mobility $Q_1(h)>0$
which is always fulfilled.
Note, that the free energy density for thermocapillarity $E_{th}$ is a function
of the ratios of viscosities $\mu$, thermal conductivities $\lambda$ and layer
thicknesses $d$. The dependence of $E_{th}$ on the viscosities, i.e.\ on
a dynamical aspect of the system, does not occur in a one-layer system.
%
%
%
\section{Results and discussion}
\label{resdis}
%
%
\subsection{Material parameters}
For our numerical investigations we focus on one specific two-layer system to
allow for a direct comparison to experiments. We chose an oil-oil system used
in Ref.\,\cite{eng}, namely silicon oil 5cS and HT70. The parameter values are
given in Tab.\,\ref{tabmat}. Note, that the given permittivities are only a 
rough estimate. 
\subsection{Linear stability}
\label{linstab}
To solve the linear problem the normal mode ansatz
\begin{equation}\label{la}
	h(x,y,t) =  h_k \exp{(ik_xx + ik_yy+\chi t)}.
\end{equation}
is used in Eq.\,(\ref{ev3d}). Linearization provides the growth rate $\chi$
\begin{eqnarray}\label{lk}
	\chi & = & -\frac{Q_1(1)}{C}\,k^2 \left(k^2 - k_c^2\right)\\
\mbox{with}\quad k_c^2 & = & C\left[ (\rho-1)G + 
\frac{\varepsilon(\varepsilon-1)^2U^2}{(\varepsilon-1+d)^3}- 3 H_1 -
\frac{3 H_2}{(d-1)^{4}}
- \frac{Q_2(1)}{Q_1(1)}
	 \frac{\lambda d M}{\left(\lambda-1 +d\right)^2}\right]
\end{eqnarray}
where $k_c$ is the cut-off or critical wavenumber and $k^2 = k_x^2 + k_y^2$. The system
is unstable for positive growth rates $\chi>0$, i.e.\ for $k<k_c$. Onset of
the instability occurs with infinite wave length when $k_c=0$.
%
\subsubsection{Gravitation, surface tension and thermocapillary effects}
\label{gsttc}
%
First, we study the situation without electric field, i.e.\ $U=0$. For
$H_1,H_2\ll 1$ the linear stability is determined by $\rho$ and $M$ only. 
In the isothermal case ($M=0$) the system is unstable for $\rho>1$, 
i.e.\ the system is  gravitationally or Rayleigh-Taylor (RT) unstable. 
In the heated case, Eq.\,(\ref{lk}) provides the critical Marangoni number
\begin{equation}
	M_c = -\frac{2}{3}\frac{(\lambda-1+d)^2(d-1)\left(\mu+d-1\right)}{\lambda d \left(\mu - (d-1)^2\right)}
	\left((1-\rho)G + 3 H_1 + \frac{3 H_2}{(d-1)^4}\right)\label{mc1}.
\end{equation}
We note that the critical Marangoni number correspond to the one found
by {\sc Smith} \cite{smith} for thin layers of viscous liquids. 
Inspection of Eq.\,(\ref{lk}) shows that the sign of the thermocapillarity term
does not only depend on the sign
of the Marangoni number but also on the sign of the mobility $Q_2(1)$.
This implies that
the sign of $Q_2(1)$ determines whether $M$ must be larger or smaller
than $M_c$ to get an instability. Denoting the zero crossing of $Q_2(h)$ by
$h_c$ (see Eq.\,(\ref{zc})), 
one finds that for $h_c>1$ the Marangoni number $M$ has to be increased
over $M_c$ for the system to become unstable, whereas for $h_c<1$ it has to be
decreased below $M_c$.
The destabilizing direction of heating is determined by substituting
$h_c=1$ in Eq.\,(\ref{zc}). Instability results if
\begin{subequations}
\begin{eqnarray}
	\mu & < &  (d-1)^2 \mbox{\hspace*{3em}} \mbox{for\hspace*{3em}}  
	M  >   M_c \\
\mbox{and if}\qquad \mu & > &  (d-1)^2 \mbox{\hspace*{3em}} \mbox{for\hspace*{3em}}  
	M  <   M_c. \label{mal1}
\end{eqnarray}
\end{subequations}
The critical system thickness (critical viscosity ratio, respectively) reads then
\begin{equation}\label{dcs}
d_c = \sqrt{\mu}+1 \mbox{\hspace{2em}}(\mu_c = (d-1)^2)
\end{equation}
as already found in Ref.\,\cite{nep1}. 
The dependence of the growth rate on the wavenumber is shown in Fig.\,\ref{fig3} for
three different situations at a system thickness $d=1.3$,
parameters from the first column in Tab.\,\ref{tabmat}, and $H_1=H_2=0$.
Because $\rho=\rho_2/\rho_1>1$ without heating the system is
Rayleigh-Taylor unstable (solid line). 
Since the system thickness is smaller than the critical one
$d_c=1.43$ (Eq.\,(\ref{dcs})), heating from below with $M>M_c=0.89$ damps
the RT instability (dashed line). As indicated in Eq.\,(\ref{mal1}) heating
from above amplifies the Rayleigh-Taylor instability (dotted line). 
The stabilization mechanism for $M>M_c$ is directly
correlated with the sign of the mobility $Q_2(h)$. The zero crossing of $Q_2(h)$ is
at $h_c\approx 0.91$ for the chosen material
parameters. Therefore, the mobility $Q_2$ is positive in the linear
regime ($h\approx 1$) and heating from below ($M>0$) acts stabilizing.

The stability diagrams
in Fig.\,\ref{fig4} show the critical Marangoni number $M_c$
(Eq.\,(\ref{mc1})) in dependence of the ratio of viscosities $\mu$ for $d=1.3$ and
$\lambda$ from the first column in Tab.\,\ref{tabmat} ($H_1=H_2=0$). The left
(right) panel represents a system that is Rayleigh-Taylor unstable (stable)
for $M=0$. At the critical viscosity $\mu_c=0.09$ the critical Marangoni number $M_c$
changes its algebraic sign in both cases. Obviously, thermocapillarity dominates
for strong heating (large $|M|$), i.e.\ RT is negligible. 

To understand the mechanism of the stabilization of a RT instability for $M>M_c$ and
$\mu>\mu_c$, we consider a small deformation of the interface in negative
$z$-direction as sketched in Fig.\,\ref{fig5}. 
First, we discuss the isothermal case ($M=0$) where for $\rho>1$ the system evolves due to
its Rayleigh-Taylor instability.
The viscous time scales of the two layers are responsible for the direction of
the fluid velocity in the vicinity of the deformation minimum.
For $d>d_c$ the viscous time scale of the lower layer $\tau_1 \propto
h_0^2/\mu_1$ is faster than the viscous time scale $\tau_2\propto
(d-h_0)^2/\mu_2$ of the upper one. Therefore, the lower layer is the {\it driving layer}
and velocities are directed away from the deformation minimum (solid arrows in
Fig.\,\ref{fig5}\,(a)). For $d<d_c$ the velocities are directed towards the deformation
(solid arrows in Fig.\,\ref{fig5}\,(b)).

When heated from below ($M>0$) the temperature is highest at the deformation
minimum. The accompanying surface
tension gradient causes a flow away from the minimum (dashed
arrows in Fig.\,\ref{fig5}). This leads to an amplification of the perturbation
for $d>d_c$, because thermocapillarity acts in the same direction as
the Rayleigh-Taylor mechanism (Fig.\,\ref{fig5}\,(a)).
For $d<d_c$ thermocapillary forces act in the same direction as before, but
the driving of the upper layer leads to flow towards the deformation minimum
(Fig.\,\ref{fig5}\,(b)). Therefore thermocapillarity damps out the Rayleigh-Taylor
instability.

Finally, we display in Fig.\,\ref{fig6} the critical Marangoni number $M_c$ in
its dependence on the system thickness $d$ as calculated from Eq.\,(\ref{mc1}).  
To avoid a destabilizing influence of a Rayleigh-Taylor instability we take the parameters
from the second column in Tab.\,\ref{tabmat}, i.e.\ we interchange the two liquids.
For $d>d_c\approx 3.34$ a minimum in $M_c$ is observed. It reflects
the antagonistic influences of the system thickness and the temperature gradient in the
lower layer ({\it driving layer}). For large $d$ the temperature gradient in the
lower layer tends to zero (from Eqs.\,(\ref{bt1}) and (\ref{tz1})), implying a
large critical Marangoni number. Decreasing the system thickness leads to
decreasing $M_c$. When decreasing $d$ further the mobility $Q_2(h=1)$ tends to
zero. Hence, for $d\to d_c$ one again finds an increasing $M_c$.   

For system thicknesses $d<d_c$ one observes a monotonously decreasing $|M_c|$ for $d\to 1$. 
Including
disjoining pressures as repelling forces changes the behavior for very small
thickness of the upper layer $d-1$ qualitatively, 
but has no influence otherwise (dashed line).  
Specifically, for $d<d_c$ the Hamaker constant of the upper layer
$H_2$ (representing the interaction of the upper substrate with liquid 1
through liquid 2) forces an extremum of $M_c$. 
Decreasing the thickness of the upper layer towards $d\approx1$ the
stabilizing Van der Waals interaction finally dominates allowing to consider 
systems with very small $d-1$ as stable.
The Hamaker constant $H_1$ (representing the interaction of the lower
substrate with liquid 2 through liquid 1) causes only a slight shift of 
$M_c$. It does not change the extremum in this region.
\subsubsection{Electrostatic field}
We conclude the discussion of the linear stability by regarding the influence of
a vertical electrical field only.
Neglecting thermocapillarity ($M=0$) and gravitation ($G=0$) yields for
$H_1,H_2\ll 1$ the critical voltage from Eq.\,(\ref{lk})
\begin{equation}\label{uc}
	U_c = \sqrt{3\left( H_1 +
\frac{H_2}{(d-1)^4}\right)\frac{(\varepsilon-1+d)^3}{\varepsilon
(\varepsilon-1)^2}}.
\end{equation}
Note, that the direction of the applied voltage ($\pm z$) has no influence on 
the stability.
Using parameters of the  second column in Tab.\,\ref{tabmat} 
with $H_1=H_2=0.01$ and $d=4$ we find the critical voltage
$U_c \approx 4.5$. A voltage of $U=30$ provides the
wavenumber $k_{max}\approx 0.18$ for the maximal growth rate $\chi$. We use
these parameters to study the time evolution with the fully nonlinear equation
below in Section~\ref{3delec}.  
\subsection{Implications of the variational formulation}
The variational formulation of the evolution equation (Eq.\,(\ref{ly1}))
provides an energy or Lyapunov functional based on a gradient energy and 
a free energy density $V(h)$ (Eq.\,(\ref{ly2})). The latter can be used in
a Maxwell construction that allows for the prediction of the character of the
resulting structure in the long-time 
evolution \cite{bes} as well as the study of metastable states \cite{thiele}.
\subsubsection{Maxwell construction}
First, we want to determine whether holes or drops are formed in the long-time
evolution. 
Since a Lyapunov functional Eq.\,(\ref{ly3}) exists, the final equilibrium thickness
profile corresponds to its global minimum. 
The mean height $h_0$ is a conserved quantity, i.e.\ an increase of the interface
height in any region is accompanied by a decrease somewhere else.
When minimizing the energy functional, this constraint has to be taken into account by a Lagrange
multiplier, $\lambda_L$, 
namely by supplementing the free energy density $V(h)$ by the term $\lambda_L h$.

Assuming a very large system in a late stage of coarsening the gradient term
of the energy functional can be neglected and the local free energy  suffices
to derive the long-time behavior.
First, consider a $V(h)$ possessing one minimum and a monotonously increasing
slope, i.e.\ a $V(h)$ with positive second derivative everywhere. Then, 
deforming the interface increases the local part of the free energy functional (which
is further increased due to the gradient
term) since the energy loss is due to mass conservation accompanied by a larger energy gain. 
Therefore, for such a $V(h)$ the (energetically minimized) final solution is a flat interface. 

Contrary to this, a $V(h)$ with two minima may allow to minimize the local 
free energy (and may even overcome the energy gain due to the gradient term)
by deforming the interface since the free energy may decrease for both $h>h_0$
and $h<h_0$. In this case the flat interface is linearly unstable and the system
will realize two film thicknesses ($h_1<h_0=1<h_2$). In analogy to spinodal
decomposition the two film thicknesses can be seen as two different phases, and
the evolution of the film thickness profile corresponds to a phase separation \cite{Mitl93}.

Mathematically formulated, the phase separation occurs if it is possible to
find a double tangent, where the curve $V(h)$ lies everywhere above this
tangent. The slope of the tangent corresponds to the Lagrange multiplier 
$\lambda_L$, and the points where it touches the curve $V(h)$ give the two
equilibrium values of $h$. In the present case the existence of the double
tangent is assured by the stabilizing disjoining pressures. Without the latter
the equilibrium film thicknesses may be found outside the gap between the two 
substrate indicating finite 'true' contact angles.
The double tangent condition is equivalent to a Maxwell construction in the
$[-d_h V(h)]$-$[h]$ space.

Resulting from mass conservation,
the ratio of the surface areas $S = S_1/S_2$ of the two equilibrium film thicknesses 
($h_1<h_0$ and $h_2>h_0$) defines the solution morphology. 
In accordance with observed structures in thin films we call $S<1$, $S>1$ and $S=1$ hole, 
drop and maze solutions, respectively.
However, for systems with $S$ close to $1$ but $S<1$ ($S>1$) the
solution reveals its visible hole (drop) character not until the final stationary state.
Note, that we use the expression 'hole' ('drop') for a hole (drop) in (of) the
lower fluid. Obviously a hole (drop) in (of) the lower fluid corresponds 
to a drop (hole) of (in) the upper fluid. 

We focus on the system thickness $d$ as control parameter for the
phase separation since $d$ can be controlled easily in experiments.
Fig.\,\ref{fig7}\,(a) shows the plot of $-d_h V(h)$ for different system thicknesses
$d$. The Maxwell point $h_M$ (via a Maxwell construction) provides then a
criterion for holes or drops. If $h_M>1$ ($S>1$) drops are preferred
(dashed line), in the other case ($S<1$) holes are expected (solid line). For $d=2$
the transition from holes to drops takes place (dotted line). We mention, that
this Maxwell point $h_{M_2}$ does not coincide with the critical system thickness
$d_c$.
\subsubsection{Metastability}
As stated in the linear investigation above in Section~\ref{gsttc},
the Rayleigh-Taylor (RT) instability can be stabilized by heating from below 
(for $d<d_c$). 
This is also confirmed by a fully nonlinear integration in time
(not shown). Nevertheless, a RT unstable system stabilized by
thermocapillarity can be metastable, i.e.\ it may be 
nonlinearly unstable to (large) finite perturbations.
This metastability can also be studied using a Maxwell
construction as shown in Fig.\,\ref{fig7}\,(b). Under isothermal conditions (dashed
line) the system is RT unstable indicated in Fig.\,\ref{fig7}\,(b) by the fact that
the vertical line at $h=1$ crosses the dashed curve in-between the two extrema.
When heating from below with $M=10$ (solid line) the system is linearly stable. However,
the Maxwell plot has still two extrema. Since the mean system thickness
($h_0=1$) is situated to the right of the maximum and the local free energy ($V(h=1)$) is larger
then the free energy of the Maxwell point ($V(h_m)$) the system is metastable.

Fig.\,\ref{fig_metas} shows the critical Marangoni number $M_c$ in
dependence of the system thickness $d$. For $d<d_c$ ($d>d_c$) heating from below (above) with
$M>M_c$ ($M<M_c$) stabilizes the system. However, a Maxwell construction shows a
metastable state for all the displayed Marangoni numbers ($|M|<60$).
This metastability was also found in experiments with oil-air layers \cite{burg}.  
A qualitative understanding of the metastability is given by the zero crossing
$h_c$ of the mobility $Q_2$ (Eq.\,(\ref{zc})). 
For perturbations larger than $|1-h_c|$ the interface is destabilized since
both gravity and thermocapillarity destabilize the system.

Fig.\,\ref{figrt_m} shows a snapshot from a 
two-dimensional numerical run for a RT instability without
thermocapillarity (dashed line). Heating from below ($M=10$) stabilizes the
system and leads to a flat stable interface for small perturbations (not
shown). However, starting with a finite perturbation in the vicinity of
$x=50$ leads to a state with a local strong modulation (solid line).
\subsection{Long time evolution}
Three-dimensional numerical 
integrations of the nonlinear equation Eq.\,(\ref{ev3d}) are done
with an ADI method (Alternating Discretization Integration). In the first half
time step the linear part is integrated implicitly in $x$-direction, in the
second half time step in $y$-direction.  The nonlinear part is calculated
explicitly. We use periodic boundary conditions in $x$ and $y$ and initially 
disturb the flat interface with small random fluctuations $\eta(x,y,t)$. Thereby the
average height is conserved ($\int{1+\eta(x,y,t)\;dx\,dy}=h_0=1$). 
Further on, we distinguish  between short-time evolution and
long-time evolution. Roughly speaking linear effects determine the dominant 
length scales of the short-time evolution. Nonlinear effects dominate the
long-time evolution that is characterized by coarsening processes.
\subsubsection{Rayleigh-Taylor instability}
In the isothermal case without electric fields, gravity is the only possibly
destabilizing influence. 
The long-time evolution of a system with $d=3>d_c$, 
$G=5$ and material parameters from the
first column in Tab.\,\ref{tabmat} is shown in Fig.\,\ref{fig_RT}. 
Initially small disturbances of the flat interface evolve into a drop structure
($t\approx 1000$). For larger times small drops vanish due to
coarsening and finally the system settles at the global energetic minimum
corresponding to one large drop (not shown). 
Fig.\,\ref{fig_RT_2} gives the evolution for $d=1.3<d_c$ and $G=20$. 
Here, the short-time evolution results in a hole pattern ($t<200$). 
Subsequently, the long-time coarsening towards structures of larger extension 
sets in ($t=1000$) and finally ends with one large hole ($t>2\times 10^5$, not shown). 

The use of $G=20$ for $d=1.3$ and not $G=5$ as for $d=3$ assures a smooth
growth in the short-time evolution. By 'smooth' we mean a gradual growth of
{\it all} holes until they have rather large amplitudes. Using instead $G=5$
for $d=1.3$ results in a rapid non-smooth hole
evolution in-between the short-time and long-time domain, i.e.\ the structure is determined by the
linear wavelength at the very beginning of the evolution only.
As soon as nonlinear terms become important ($|\eta(x,y.t)|\ll 1$ is violated)
only part of the linearly developed structure evolves.
Here, this rapid hole evolution is caused by
the stabilizing mechanism of the disjoining pressure at the upper plate
(Hamaker constant $H_2=0.01$) which can no longer be neglected even for small
perturbations of the flat interface. This affects both the linear and the nonlinear
evolution of the interface. Namely, it suppresses interface evolutions for $h>1$
and therefore causes a rapid evolution for $h<1$. 
We note again, that here the rapid hole evolution is caused by disjoining pressures, whereas the mobility $Q_1(h)$ has no effect.

Fig.\,\ref{fig_RT_M} displays the time evolution for $d=2$, $G=5$, 
and parameters from the first column of Tab.\,\ref{tabmat}. For these parameters,
neither drops nor holes are energetically preferred.
This leads to a clearly visible maze or labyrinth structure 
which also shows the typical coarsening dynamics at long times.
All long-time solutions shown (drops for $d=3$, holes for $d=1.3$ and maze
structures for $d=2$) correspond to the predictions of the Maxwell
construction in Fig.\,\ref{fig7}. 

To quantify the coarsening behavior we calculate at each timestep 
the mean wave number
\begin{equation}\label{km}
	\langle k \rangle\,(t) =
\frac{{\displaystyle\int_{k_x}\!\int_{k_y}}{dk_x\,dk_y\;\sqrt{\vec k^2}\;\tilde
h^2(k_x,k_y,t)}}{{\displaystyle\int_{k_x}\!\int_{k_y}}{dk_x\,dk_y\;\tilde
h^2(k_x,k_y,t)}}
\end{equation}
where $\tilde h(k_x,k_y,t)$ denotes the Fourier transform of $h(x,y,t)$.
We mention, that the wavevectors are distributed
on a small annulus. Therefore the approximation
$\langle k\rangle^2\approx \langle k^2\rangle$ holds 
and the mean wave number $\langle k\rangle$ can also be taken as a
qualitative measure of the mean curvature of $h(x,y,t)$. 
Fig.\,\ref{fig_RT_K} shows the dependence of $\langle k\rangle$ and of the
corresponding energy (Eq.\,(\ref{ly3})) on time for the three numerical
evolutions discussed above. 
The energy decreases always monotonously in time as expected. 
The mean wave number $\langle k\rangle$ shows two local extrema at 
$t_{min}$ and $t_{max}$, respectively, with $t_{min} <t_{max}$.  
In the region between the two extrema the amplitude of the interface deflections
outgrows the linear regime. Further on, the averaging in Eq.\,(\ref{km}) allows
to interpret the strength of the maximum as a measurement for global amplitude
growth. The absence of a local maximum indicates that
all linearly evolved drops or holes evolve globally and uniformly towards larger
amplitudes, whereas a strong peak indicates a nonlinear evolution of a few linearly
evolved drops (holes) only.

The mean wave number at the local minimum corresponds to the wave number of
the maximal growth rate from the linear investigation (thin horizontal
lines). Hence the evolution for $t<t_{min}$ is determined by linear terms,
i.e.\ the wavenumber with the maximal linear growth rate emerges in the
system. Therefore, the region around the two extrema can be considered as the
frontier between short-time and long-time evolution. 
In the long-time evolution ($t>t_{max}$) nonlinear effects dominate and a 
scaling law \cite{bes} 
\begin{equation}\label{kt}
	\langle k\rangle = c\cdot t^{-\beta}
\end{equation}
can be extracted which reflects the coarsening of the system for long times. 
To determine a 'true' scaling exponent a
statistically significant average of many numercial
runs with different initial conditions is necessary.
Due to the strongly time-consuming character of the necessary
computer calculations only a few runs were used to determine
the respective tendencies of scaling exponents presented here.
However, in the following we call them shortly 'scaling exponents'.

We find a nearly identical scaling exponent of $\beta\approx 0.14$ for drops
($d=3$, solid line, $t_{max}\approx 150$), holes ($d=1.3$, dashed line, $t_{max}\approx
210$) and maze structures ($d=2$, dotted line, $t_{max}\approx 320$).   
Neither the system thickness $d$ nor the gravity number affect the
long-time scaling.  

To identify the acting coarsening modes we illustrate the flow pattern
by calculating differences between the interfaces $h(x,y,t_1)$ and $h(x,y,t_2)$ 
at different times $t_1$ and $t_2$, respectively, shown for the evolution of
drops in Fig.\,\ref{fig_RT_diff}  (corresponding to Fig.\,\ref{fig_RT}). 
One can identify two different coarsening mechanisms being dominant at different times
{\it within} a single long-time evolution. 
As a result of the short-time evolution many small drops exist. Neighboring drops
attract each other strong enough to move the entire (small) drops and 
finally combine to one large drop sitting at an intermediate position.
This translational coarsening mode is illustrated in Fig.\,\ref{fig_RT_diff}\,(a)
where its signature in the difference plot is that all drops have white (mass
gain) and black (mass loss) parts of their edges.

For larger times a transition from the dominant
translational mode to a dominance of the volume transfer mode takes places. Now the mean
distance of the drops is too large to get the (large) drops moving. Only material is
transported between the sitting drops resulting in a slow disappearance of
smaller drops and the growth of the larger drops. This mass transfer mode 
is illustrated in Fig.\,\ref{fig_RT_diff}\,(b)
where its signature in the difference plot is that there exist drops that have
completely white or completely black edges. 

Finally, we want to show that our descriptive explanation for the viscous
timescales (and therefore for the interface velocity directions) given above
in Section~\ref{linstab} is in
concordance with the fully nonlinear evolution. We integrated Eq.\,(\ref{ev})
numerically in two dimensions for a RT unstable system with $H_1=H_2=0$ (integration was
stopped before film rupture occurred).  
The $x$-components of the velocities $u_1(x,z,t)$ and $u_2(x,z,t)$  can be calculated from
Eq.\,(\ref{velu}). They are plotted in Fig.\,\ref{figrt_vel} (right plots) for
the position $x=20$.  
The left panels of Fig.\,\ref{figrt_vel} display the interfaces and the contour lines of
the streamfunction $\varphi$ ($u = -\partial_z\varphi$, solid lines
$\varphi>0$, dashed lines $\varphi<0$). 
For $\mu<\mu_c$ (Fig.\,\ref{figrt_vel}\,(a)) the interface velocity is negative, i.e.\
fluid moves away from the deformation minimum and therefore the lower layer is
the {\it driving layer} (solid arrows in Fig.\,\ref{fig5}\,(a)). For $\mu>\mu_c$
(Fig.\,\ref{figrt_vel}\,(b)) the interface velocity is positive. Hence fluid moves to
the deformation minimum and the upper layer is the {\it driving layer} (see also
solid arrows in Fig.\,\ref{fig5}\,(b)). 
\subsubsection{Thermocapillary effects}
In this section we include thermocapillary effects (using
Eq.\,(\ref{stressb})). Thermocapillarity can act both stabilizing and
destabilizing as seen from linear analysis. To avoid an amplification due to
gravity we use parameters from the second column in
Tab.\,\ref{tabmat}. Therefore, $\rho<1$ and gravity stabilizes the flat film. 
Eq.\,(\ref{mal1}) provides then the critical system thickness $d_c \approx
3.34$. An unstable initial flat film is obtained for $d>d_c$ ($d<d_c$) 
by heating from below (above). In the following,
we use $d=2$ and $d=4$ to illustrate the destabilization by different
directions of heating. 

For $d=2$, the critical Marangoni number is $M_c \approx -5$ and the system
becomes unstable for $M<M_c$. Fig.\,\ref{fig_M_1} shows a numerical run for
$M=-10$.  Small initial disturbances evolve smoothly into drops which coarse
in the long-time evolution ($t=3\times 10^5$)  corresponding to the
prediction of a Maxwell construction (Maxwell point $h_M\approx 1.34$). 
Fig.\,\ref{fig_RT_K_1} (solid line) shows the mean wave number $\langle
k\rangle(t)$ and the energy versus time. The energy is again a  monotonously
decreasing function in time. 
The mean wave number $\langle k\rangle(t)$ has again two local extrema. The
minimum reached at $t_{min}\approx 1800$ corresponds to the fastest linear
wave number.
The long-time coarsening sets in after the local maximum at $t_{max}\approx 4000$
and the scaling coefficient defined in Eq.\,(\ref{kt}) is determined to be
$\beta\approx 0.16$. 

For $d=4$ the critical Marangoni number is $M_c \approx 37.84$ and we use
$M=70$ for the numerical run displayed in Fig.\,\ref{fig_M}.  
Starting from small perturbations one hole evolves rapidly
at $t\approx 700$. Subsequently, more and more holes arise ($t=1100$). For
$t>1100$ coarsening sets in and in the long-time limit a single hole
remains ($t=2\times 10^5$). This corresponds to the prediction of the Maxwell
construction (Maxwell point $h_M\approx 0.51$).
The mean wave number $\langle k\rangle$ in Fig.\,\ref{fig_RT_K_1} (dashed line)
shows again a minimum  corresponding 
to the fastest linear wave number and a very pronounced maximum 
 indicating the rapid evolution of one hole
between the two extrema.
Since the averaging in Eq.\,(\ref{km}) gives approximately also the root of the mean curvature 
the abrupt rise of $\langle k\rangle$ is obvious even for the evolution of a single hole.
The long-time scaling is with $\beta\approx 0.27$ remarkably faster than for $d=2$. 
Note, that we found numerically that the scaling exponent for $d=4$ does
nearly not depend on the Marangoni number. 

The differences in the short-time evolution for $d=2$ and $d=4$
(rapid evolution of one hole for $d=4$ versus
smooth evolution of many drops for $d=2$) can be understood in terms of the
effective mobilities
$Q'_1(h)=G(1-\rho)\,Q_1(h)$ and $Q'_2(h) = M\,Q_2(h)$. shown in 
Fig.\,\ref{fig_MOB}. For
$d=4$, $Q'_2$ crosses zero close to $h=1$ (thick lines, $h_c\approx 1.2$ from
Eq.\,(\ref{zc})) and $Q'_1$ increases for increasing $h$. 
Therefore, the interface evolution is slowed down for $h>1$ (and finally stopped for
$h>h_c$) and accelerated for $h<1$. This results in a rapid hole evolution.
For $d=2$,  both mobilities show an approximate
symmetry around $h=1$ (thin lines). Hence, no interface thickness is suppressed 
allowing for a smooth evolution of drops.   
\subsubsection{Application of an electric field}
\label{3delec}
Finally, we illustrate the time evolution caused by a vertically applied
electrical field. We use the parameters from the second column in 
Tab.\,\ref{tabmat} for an isothermal ($M=0$) system with $G=0$. 
Fig.\,\ref{fig_UC} shows snapshots of the long-time evolution for an applied
voltage $U=30$. Initially small disturbances of the interface evolve smoothly
to drops and the long-time coarsening sets in at $t\approx 10000$. The 
dependence of the mean wave number on time shown as dotted line in
Fig.\,\ref{fig_RT_K_1} shows a minimum at $t_{min}\approx 1500$ and a maximum at
$t_{max}\approx 4700$. Again, the minimum coincidences to the wave number of
the fastest linear mode $k_{max}\approx 0.18$.

The derived long-time scaling exponent $\beta\approx 0.04$ is small compared
to the exponents measured above for the Rayleigh-Taylor and thermocapillary
instabilities. In absolute values we find only a small change from 
$\langle k\rangle=0.18$ at $t=10^4$ to $\langle k\rangle=0.16$ at $t=10^5$. 
We conclude, that one can consider the length scale of the pattern 
in the long-time evolution as frozen to the value of the wavelength of the
fastest linear mode ($2\pi/k_{max}$) at least up to $t=10^5$.    
%
%
%
\section{Conclusion}
\label{summ}
%
%
Using long-wave approximation we have derived a single evolution equation for
the interface profile of a two-layer system bounded by rigid plates. 
This equation is written in a general form to facilitate the inclusion of
arbitrary body forces and normal or tangential forces at the interface. 
In the analysis of the model presented here, we have focused on the
influences of gravity, thermocapillarity and electrostatic fields. 

We have shown that the mobility of the normal-stress and the body force terms
$Q_1(h)$ is always positive as in the one-layer case. However, it tends to
zero not only for $h\rightarrow 0$ as for the one-layer system
but also for $h\rightarrow d$. The latter has no counterpart in one-layer
systems where the mobility increases monotonously with the film thickness
\cite{oron1}. The second qualitative difference is the sign change
of the mobility for the shear-stress term $Q_2(h)$. 
Both mobilities
can affect strongly the linear and nonlinear evolution. 
For instance, the direction of heating needed for destabilization is determined
by the zero crossing of the mobility $Q_2(h)$.
Furthermore, the shapes, zeros and extrema of the mobilities allow at least a
qualitative prediction of the dynamics of the system without any numerical
investigation.  
Moreover, in contrast to weakly nonlinear theories \cite{nep1} we are able to check these
descriptive criterions integrating  the fully nonlinear 
equation numerically. A non-smooth rapid hole (drop) evolution in-between the short-time
and long-time regime can already be estimated from the trend of the mobilities. 

Remarkably, although in the heated case 
the system is dissipating energy through convection within the drops (or around
the holes) even when the final
stable state is reached, the use of long-wave approximation allows
for a variational formulation using an energy or Lyapunov
functional for the film thickness profile. The film thickness
evolution equation takes then the form of the simplest possible
equation for the dynamics of a conserved order parameter field 
\cite{Lang92,Mitl93,OrRo92}. A prominent representative of this class of
systems is the Cahn-Hilliard equation describing the
evolution of a concentration field for a binary mixture \cite{Cahn65}. 
In contrast to the one-layer case \cite{OrRo92,thiele}, here the energy
itself depends on material parameters that characterize the dynamics 
of the system, namely the ratio of viscosities.
We used the energy functional to predict the expected long-time 
behavior, namely the evolution of holes, drops
or maze structures. It also allows for the study of  metastable states.
The predictions have been confirmed by fully
nonlinear numerical integrations of the evolution equation.

Using a linear stability analysis we have discussed the conditions for a
gravitational or Rayleigh-Taylor instability, thermocapillary destabilization
and stabilization, and an electrohydrodynamic instability for dielectric
liquids under the influence of an electrical field.
We have shown that thermocapillarity may act stabilizing as well as
destabilizing depending on material parameters.
The behavior becomes intuitively clear because when treating both 
layers in the same way no direction of heating should be preferred. This implies that 
depending on material parameters both ways to destabilize the system --
heating from above and heating from below -- have to be possible. 

We have given special emphasis to the study
of the possibility to stabilize a Rayleigh-Taylor unstable two-layer system
by heating from below. This seemingly counter-intuitive behavior first 
discussed in Ref.\,\cite{nep1} is a typical property of multi-layer systems 
and is directly correlated with the change of sign of the mobility $Q_2(h)$.
However, we have shown that the stabilized Rayleigh-Taylor system is metastable.
This explains a problem encountered in the experiments of 
{\sc Burgess et al.} \cite{burg}. Although they could stabilize a
Rayleigh-Taylor unstable 'oil on air' system by heating from below this was
only possible in ten percent of the experimental runs. This is due to the fact
that the preparation of the initial flat film involved large amplitude
disturbances. Because of the metastability of the system this results in
the destabilization of ninety percent of the runs because the disturbance is
larger than the critical one.
The transition from the two- to the three-dimensional equation has shown that
a weak mean flow arises. We neglect this additional flow in the main part of
the present work due to its very weak influence on the presented results.
However, the mean flow has been discussed in detail in the appendix.

We have implemented numerical schemes for two- and three-dimensional
versions of the fully nonlinear equation and have given an overview showing
different possible long-time evolutions consisting of coarsening  
hole, drop or maze patterns. Furthermore, we have analysed the
scaling behavior by calculating the time dependence of the mean wave number
of the patterns and extraction of 
the tendency of scaling exponents. 

Isothermal two-layer systems, i.e.\ taking into account 
gravitation and disjoining pressures only, show 
the same long-time scaling for different system thicknesses $d$. 
However, incorporating thermocapillarity the system thickness $d$ affects 
the long-time scaling essentially. 
This scaling behavior is in contrast to the one-layer case which was found to be determined
by {\it one} scaling exponent  ($\beta\approx 0.21$) \cite{bes}.
The one-layer coefficient lies within the two-layer range $0.16\le \beta \le
0.27$ found here. 
Moreover, an isothermal ($M=0$) and agravic ($G=0$) electrohydrodynamically unstable 
system reveals a very small scaling coefficient ($\beta\approx 0.04$). To our 
knowledge for this class of evolution equations such a slow long-time scaling is
found for the first time. 

Finally, we have shown that tangential interface forces (thermocapillary
forces) allow for rapid hole (or drop) formation after the short-time
evolution. Again, this mechanism can be understood in terms of the change of
sign of the tangential mobility $Q_2(h)$.    

In the present work,
disjoining pressures were solely used to inhibit rupture of the layers to allow for 
the study of the long-time behavior. This corresponds to the assumption that 
liquid 1 and 2 wet the lower and upper plate, respectively. However, our
present theory is not apt to describe situations where both
layers are ultrathin (less than $100\,nm$), a situation gaining more and more 
importance for research communities and industrial applications. Then 
disjoining pressures dominate and the used terms are not exact enough
because part of the forces between the lower (upper) fluid and the upper
(lower) substrate are neglected. Furthermore, the used Van der Waals
interaction should be supplemented by additional short-range interactions
\cite{Isra92}. A further analysis of
disjoining pressures in two-layer systems seems promising and will be
published elsewhere.  

We emphasize our results for the action of a vertical electric field
since recent experiments focus on such systems, as for example, done by  
{\sc Lin et al.} using two polymeric liquids \cite{lin2,lin}.  
They monitored the time evolution up to the impingement of the lower polymer layer
on the upper electrode and showed a series of snapshots of the evolving
morphology (Fig.\,4 of Ref.\cite{lin2}).  Interestingly, they 
found a nearly constant length scale of the evolving columnar structures from
the early stage on, corresponding to the fastest linear mode.
This corresponds to the results of our linear and nonlinear analysis of this case
that we performed for a comparable ratio of permittivities. Furthermore, a visible
concordance of both timeseries exists. The small scaling exponent $\beta\approx 0.04$
found here can be regarded as a structure length frozen to the fastest linear
wavelength. This indicates, that our model gives reasonable results even for macromolecular
liquids. 

Finally, we stress the advantages for physics as well as industrial applications of 
bounded two-layer systems. 
From a physical point of view our single interface equation captures both the
long-wave evolution and the interface interactions of two fluids. Therefore it
allows for detailed investigations how the fluid properties of both the upper and 
the lower fluid layer determine the stability, metastability and short-time as well as 
long-time evolution.
Furthermore ultrathin films play already a major role to create desired structures or 
stable flat interfaces.
Usually one-layer equations are used for that kind of industrial applications. 
However, controlled boundary conditions, well defined bulk properties and 
consequently well defined interface actions enhance the accuracy of experiments
as well as the examination of theory and experiment.

\appendix*
\section{Three-dimensional evolution equation}
The generalization of the derivation of the film thickness equation to
three dimensions is straightforward up to
Eq.\,(\ref{int}) for which the three-dimensional (already integrated) 
version reads
\begin{equation}\label{int3d}
	\nabla\cdot\left[A(h)\left(\nabla \tilde P_1 - F_1(h)\nabla({\cal N}+\Phi) 
	-F_2(h) \vec{\cal T}\right)\right] =0
\end{equation}
where $\nabla=(\partial_x,\partial_y)$ and
\begin{equation}
	A(h) = -\frac{D}{12\mu((\mu-1)h+d)}
\end{equation}
with $D$ from Eq.\,(\ref{cf}).

Eq.\,(\ref{int3d}) cannot be solved analytically for the pressure $\tilde P_1$. 
However, to
fulfill Eq.\,(\ref{int3d}) its argument must be of the form
\begin{equation}\label{fstream}
	\frac{1}{A(h)}\,\mbox{\bf rot}(f\,\vec e_z)
	= \nabla \tilde P_1 - F_1(h)\nabla({\cal N}+\Phi) 
	-F_2(h) \vec{\cal T}
\end{equation}
with $f$ as an arbitrary scalar function, $\mbox{\bf rot}(f\,\vec e_z)=(\partial_yf,-\partial_xf,0)$ and $F_1(h)$ from Eq.\,(\ref{f1}).
Note that this function $f$ is already present without symmetry breaking effects (e.g.
inclined or rotating systems).  
Substitution of the gradient pressure from Eq.\,(\ref{fstream}) in the three
dimensional correspondent to Eq.\.(\ref{ev1}) 
 provides the evolution equation for $h$
\begin{subequations}\label{meaneq}
\begin{eqnarray}
	\partial_t h  &  = &
	\nabla\cdot\left[ Q_1(h)\,\nabla( {\cal N} + \Phi)
	+Q_2(h)\,\vec{\cal T}\right] + \nabla\cdot\left[Q_3(h)\,\mbox{\bf rot}(f\,\vec e_z) \right]
	\label{meanev}
\end{eqnarray}
where $f$ as a function of $h$ and its spatial derivatives 
is given implicitly by
\begin{eqnarray}	
	-\frac{1}{A(h)}\triangle f  & = & 
	\partial_h\left(\frac{1}{A(h)}\right)\mbox{\bf rot}(h\,\vec
e_z)\cdot\mbox{\bf rot}(f\,\vec e_z)
	+\partial_h F_1(h)\,\mbox{\bf rot}(h\,\vec e_z)\cdot\nabla({\cal N}+\Phi) 
	\nonumber\\
	& & +\partial_h F_2(h)\,\mbox{\bf rot}(h\,\vec e_z)\cdot\vec{\cal T}
	+ F_2(h)\,\mbox{\bf rot}(\vec{\cal T})\cdot\vec e_z,\label{meanflow}
\end{eqnarray}
\end{subequations}
resulting from applying the $\mbox{curl}$ to 
Eq.\,(\ref{fstream}).

The third mobility
\begin{equation}
	Q_3(h) = F_1(h)-1 
\end{equation}
is a monotonically decreasing function in $h$ ($Q_3(0)=0, Q_3(d)=-1$).
Note that $\partial_h Q_3(h)=\partial_h F_1(h)$ is zero for $h=0$ and $h=d$ and negative
for $0<h<d$. 
The additional function $f$ reflects the possible mean flow of the system
induced by a vertical vorticity contribution.
We mention that in the one-layer limit ($\mu\to 0$ or $d\to\infty$) this 
vorticity contribution is zero ($1/A(h)\to 0$) and the usual normal
condition for the pressure is recovered.

Substituting $\cal N$, $\Phi$ and $\cal T$ (from subsection \ref{effect}) in the rhs of
Eq.\,(\ref{meanflow}) provides
\begin{equation}\label{flowour}
	\frac{1}{A(h)}\triangle f 
	+ \partial_h\left(\frac{1}{A(h)}\right)\mbox{\bf rot}(h\,\vec
e_z)\cdot\mbox{\bf rot}(f\,\vec e_z)
	=
	C^{-1}\partial_h F_1(h)\,\mbox{\bf rot}(h\,\vec e_z)\cdot\nabla(\triangle h) 
\end{equation}
showing that solely the surface tension contributes to $f$.

Due to the properties of $\partial_h F_1(h)$ the vorticity $f$ is basically determined by the geometrical shape of the interface.
It can be shown with Eq.\,(\ref{flowour}) that rotationally symmetric structures are not
affected in the strongly nonlinear regime by this additional vorticity contribution.

Furthermore Eq.\,(\ref{flowour}) reveals that 
for small deviations from symmetric states the vorticity $f$ acts in 
the same direction as surface tension itself in  Eq.\,(\ref{meanev}).

We have checked and confirmed all numerical runs of this paper taking
into account Eq.\,(\ref{flowour}).
In general the contribution of $f$ is small. Therefore we neglect it and
consider the evolution equation (\ref{ev3d}) only.
However, we note that the effects of Eq.\,({\ref{flowour}) on the evolution
equation are an
attractive and important subject for future investigations.

\newpage
\begin{table}[ht]
\begin{tabular}{|l|r|r|}
	\hline
	Fluid 2 - Fluid 1\hspace{2em} &\hspace{1em}  HT70 - silicon oil 5cS &\hspace{1em} silicon oil 5cS - HT70 \\
	\hline	
 	density $\rho=\rho_2/\rho_1$  & $1.826$ & $0.548$ \\
	\hline
 	viscosity $ \mu = \mu_2/\mu_1$  & $0.1826$ & $5.48$ \\
	\hline
 	thermal conductivity $\lambda=\lambda_2/\lambda_1$  \hspace{2em}  & $0.598$ & $1.67$ 
	\\
	\hline
 	permittivity $\varepsilon=\varepsilon_2/\varepsilon_1$    &$0.77$ & $1.3$\\
	\hline
\end{tabular}
\caption{Material parameters for a silicon oil 5cS - HT70 system taken from
Ref.\,\cite{eng}. The values in the first column are for HT70 on silicon oil 5cS,
whereas for the second column the liquids are interchanged, i.e.\ silicon oil 5cS on
HT70.}\label{tabmat}
\end{table}

\newpage
Figure captions:
\begin{itemize}
\item[Fig.\,\ref{fig1}:] Sketch of the system. The two layered immiscible liquids are bounded by two rigid smooth plates. The flat interface is 
situated at the mean height $h_0$. The position $h(x,y,t)$ of an evolving interface profile is a function of $x$, $y$ and $t$ only.

\item[Fig.\,\ref{fig2}:]
Shown are the mobilities $Q_1(h)$ (normal-stress and body force terms)
and $Q_2(h)$ (tangential-stress term) for $d=1.3$ and  $\mu=0.1826$. The
mobility $Q_1(h)$ is always positive and vanishes for $h\to 0$ and $h\to d$. 
The mobility $Q_2(h)$ changes sign at $h_c\approx 0.91$.

\item[Fig.\,\ref{fig3}:]
The dependence of the linear growth rate $\chi$ on wavenumber $k$ for $d=1.3$,
$\rho=1.826$, $\mu=0.1826$ and $\lambda =0.598$ (parameters from the first column
in Tab.\,\ref{tabmat}) at $G=1$ and $C^{-1}=20$ ($H_1=H_2=0$). In the
non-isothermal case $M=\pm 1$.

\item[Fig.\,\ref{fig4}:]
The stability diagrams show the critical Marangoni number $M_c$ in
dependence of the ratio of viscosities $\mu$ for $d=1.3$, $\lambda=0.598$ and 
$G=5$ ($H_1=H_2=0$). $M_c>0$ ($M_c<0$) corresponds to heating from below (above).
The direction of heating leading to destabilization changes 
at the critical viscosity $\mu_c=0.09$. The 
left (right) panel corresponds for $M=0$ to a Rayleigh-Taylor unstable (stable) system 
with $\rho=1.826>1$ ($\rho=0.548<1$).

\item[Fig.\,\ref{fig5}:]
Sketch illustrating the mechanism of the (a) destabilizing and (b)
stabilizing thermocapillary action for a Rayleigh-Taylor (RT) unstable system
heated from below. Solid arrows indicate the
liquid velocity close to the interface in the respective {\it driving layer} 
for pure RT, dashed arrows signal the effect of
thermocapillarity. (a) Due to thermocapillarity the total flow in the {\it driving
layer} is directed away from the deformation as known from one-layer systems. 
This causes an amplification of the disturbance.
(b) The viscous timescale $\tau_{\mu_2}\propto (d-h_0)^2/\mu_2$ of the upper
layer is faster than the one of the lower layer. The influence of the upper
{\it driving layer} dominates leading to flow to the deformation minimum. 
Thermocapillarity causes flow in the opposite direction, thereby weakening or
completely damping the disturbance.

\item[Fig.\,\ref{fig6}:]
Shown is the critical Marangoni number $M_c$ versus the system thickness
$d$ for $\rho=0.548$, $\mu=5.48$, $\lambda = 1.67$ and $G=5$. The critical
system thickness is $d_c\approx 3.34$. For $d<d_c$ heating 
from above acts destabilizing and $M_c$ is a monotonic function of
$d$. For $d>d_c$ heating from below destabilizes and a minimum results from
competing mechanisms (see main text). Inclusion of stabilizing disjoining pressures cause a
maximum of $M_c$ for $d<d_c$ but have nearly no influence for $d>d_c$ 
(dashed lines).

\item[Fig.\,\ref{fig7}:]
Maxwell constructions (horizontal lines) 
based on the local energy are given for
(a) three different system thicknesses $d$ (see legend) of a Rayleigh-Taylor unstable
system at $G=5$, $C^{-1}=20$, $H_1=H_2=0.01$, $M=0$, $\rho=1.826$ and
$\mu=0.1826$. For $d=2$, the Maxwell point is $1$ (vertical dotted
line). $d>2$ leads to drop solutions (dashed line). Hole solutions are
expected for  $d<2$ (solid line). (b) Illustrates the occurrence of metastable states
using $G=10$, $H_1=0.1$, $H_2=0.01$, $\rho=1.826$, and $\mu=0.1826$, 
i.e.\ an Rayleigh-Taylor unstable system. The isothermal case (dashed line) is
linearly unstable, whereas heating from below ($M=10$, solid line) stabilizes the system
linearly. However, the local maximum still exists at $h<1$ indicating
a metastable flat film.

\item[Fig.\,\ref{fig_metas}:]
Shown is the critical Marangoni number $M_c$ versus system thickness $d$ for
$G=10$, $C^{-1}=20$, $\rho=1.826$, $\mu=0.1826$, $\lambda=0.598$, $H_1=0.1$
and $H_2=0.01$. The isothermal system is Rayleigh-Taylor unstable ($M=0$). 
For $d>d_c$ ($d<d_c$) heating from above (below) stabilizes the system. However, for the shown range of $M$ the system
remains metastable, i.e.\ finite 
disturbances larger than a critical nucleus will grow.

\item[Fig.\,\ref{figrt_m}:]
Snapshots from of the nonlinear evolution of the interface for $d=1.3$, $G=10$,
$C^{-1}=20$, $\Delta t=0.1$, $\Delta x =0.5$, $\rho=1.826$, $\mu=0.1826$,
$\lambda=0.598$, $H_1=0.1$ and $H_2=0.01$. Without heating ($M=0$) the system
is linearly  Rayleigh-Taylor unstable implying the growth of  infinitely 
small disturbances (dashed line). 
For $M>M_c = 4.62$ the system is metastable. We applied for $M=10>M_c$ a strong disturbance of the
interface ($h\approx 0.7 \pm 0.1$) in the vicinity of $x=50$.

\item[Fig.\,\ref{fig_RT}:]
Three-dimensional 
snapshots of the long-time evolution of a Rayleigh-Taylor unstable system
at $d=3$, $G=5$, $C^{-1}=20$,
$\rho=1.826$, $\mu=0.1826$ and $H_1=H_2=0.01$. The system size is
$L_x=L_y=200$ with a resolution $\Delta t=0.1$, $\Delta x =\Delta y =2$. Initial (small) perturbations of the flat interface lead to
drop formation, and subsequent long-time coarsening.  
Finally, one single drop survives (not shown).

\item[Fig.\,\ref{fig_RT_2}:]
Three-dimensional snapshots of the long-time evolution of a
Rayleigh-Taylor unstable system  
at $d=1.3$, $G=20$, $C^{-1}=20$, $\rho=1.826$, $\mu=0.1826$ and $H_1=H_2=0.01$. The system size  is
$L_x=L_y=100$ with a resolution $\Delta t=0.1$, $\Delta x =\Delta y
=0.5$. From initially small perturbations holes start to evolve
($t=150$) smoothly. Subsequently long-time coarsening sets in at $t\approx
1000$ and finally an one-hole solution is reached
($t>2\times10^5$).

\item[Fig.\,\ref{fig_RT_M}:]
Three-dimensional snapshots of the long-time evolution of a maze
structure in a Rayleigh-Taylor unstable system at
$d=2$, $G=5$, $C^{-1}=20$, $\rho=1.826$, $\mu=0.1826$ and $H_1=H_2=0.01$. The system size 
is $L_x=L_y=200$ with a resolution $\Delta t=0.1$, $\Delta x
=\Delta y =2$. It is clearly visible that neither holes nor drops are
energetically preferred. In the long-time evolution ($t>1000$) the usual
coarsening takes places.

\item[Fig.\,\ref{fig_RT_K}:]
Shown are (a) the mean wavenumber $\langle k\rangle$  and (b) the
energy $E$ in dependence on time for Rayleigh-Taylor unstable systems with 
$\rho=1.812$, $\mu=0.1826$, $H_1=H_2=0.01$, and different thicknesses $d$ as
given in the legend. Horizontal thin lines give the corresponding fastest linear wave
numbers.

\item[Fig.\,\ref{fig_RT_diff}:]
Grey-level plots of the interface height $h$ 
at two timesteps and the difference of the two images
for the time evolution in Fig.\,\ref{fig_RT}. In the difference plot dark
(light) areas indicate mass loss (gain) of the lower layer
(a) During the first stage of the long-time evolution
neighboring drops move towards each other to merge indicating the dominance
of the translational mode of coarsening. (b)
At a later stage, small drops shrink and neighboring
large drops grow, indicating the dominance of the mass transfer mode of coarsening.

\item[Fig.\,\ref{figrt_vel}:]
Given are (left) film thickness profiles $h(x,t)$  and (right) 
$x$-components of the
fluid velocity $u(x=20,z,t)$ for $G=5$, $C^{-1}=20$, $\rho=1.826$, $\mu=0.1826$, $H_1=H_2=0$
and a resolution $\Delta t=0.1$,
$\Delta x = 0.5$. The left panels also
show contour lines of the streamfunction $\varphi$ 
($u = -\partial_z\varphi$).
(a) In a system where $\mu < \mu_c$ ($d=3$) at $t=120$ the interface height 
$h(20,t)<1$ and $\partial_x h(20,t)<0$ provides  $u<0$ at
the interface. The direction of the interface velocity corresponds to the solid
arrows in Fig.\,\ref{fig5}\,(a), i.e.\ the lower layer is the {\it driving layer}. (b)
If $\mu>\mu_c$ ($d=1.3$), at $t=1200$ the
interface velocity $u$ at $x=20$ is positive and the upper layer is the {\it driving layer} (compare to solid arrows in Fig.\,\ref{fig5}\,(b)).

\item[Fig.\,\ref{fig_M_1}:]
Three-dimensional snapshots of the long-time evolution of a Marangoni instability for
$M=-10$, $d=2$, $G=5$, $C^{-1}=20$, $\rho=0.548$, $\mu=5.48$, $\lambda=1.671$, $H_1=0.01$ and
$H_2=0.05$. The system size is $L_x=L_y=100$ with a resolution  $\Delta t=0.1$, $\Delta x
=\Delta y =1$. From initially small
perturbations the systems evolves smoothly to a single drop solution in the
long-time limit ($t>10^6$, not shown).

\item[Fig.\,\ref{fig_RT_K_1}:]
Shown are (a) the mean wavenumber $\langle k\rangle$  and (b) the energy $E$
in dependence on time for two thermocapillary and one electrohydrodynamic
unstable system (see legend) with $\rho=0.548$, $\mu=5.48$ and
$\lambda=1.671$. 
Horizontal thin lines give the corresponding fastest linear wave numbers.

\item[Fig.\,\ref{fig_M}:]
Three-dimensional snapshots of the long-time evolution of a Marangoni
instability for $M=70$, $d=4$, $G=5$, $C^{-1}=20$,  $\rho=0.548$, $\mu=5.48$, $\lambda=1.671$, $H_1=0.05$ and
$H_2=0.01$. The system size is $L_x=L_y=200$ with a resolution $\Delta t=0.1$, $\Delta x
=\Delta y =1$. We started with initially small
perturbations. At $t\approx 700$ one hole starts to evolve rapidly and
subsequently more and more holes arise. At $t\approx 1100$ long-time
coarsening sets in and continues until a single large hole is reached
($t>2\times 10^5$).

\item[Fig.\,\ref{fig_MOB}:]
Shown are the rescaled mobilities $Q'_1=G(1-\rho)\,Q_1$ and $Q'_2=
M\, Q_2$ for thermocapillary unstable systems  with $d=2, M=70$ (thin lines,
numerical run in Fig.\,\ref{fig_M_1}) and  $d=4, M=-10$ (thick lines, numerical
run in Fig.\,\ref{fig_M}). 
For $d=4$ the zero crossing of $Q'_2$ is at $h\approx 1.2$. This leads to a
suppressed interface evolution for $h>1$ resulting in a rapid hole evolution.
For $d=2$ smooth drop evolution results since the regions $h>1$ and
$h<1$ are roughly symmetric.

\item[Fig.\,\ref{fig_UC}:]
Three-dimensional snapshots of the long-time evolution of an
electrohydrodynamic instability for $d=4$, $C^{-1}=20$, $U=30$,
$\varepsilon=1.3$, $\mu=5.48$ and
$H_1=H_2=0.01$. The system size is $L_x=L_y=300$ with a resolution
$\Delta t=0.1$, $\Delta x =\Delta y =3$. One finds a smooth short-time
evolution of drops. The long-time coarsening sets in at $t\approx
10000$ and the long-time scaling exponent is very small. 
\end{itemize}

\clearpage
\begin{figure}[ht]
\begin{center}
\includegraphics[angle=0,width=\textwidth]{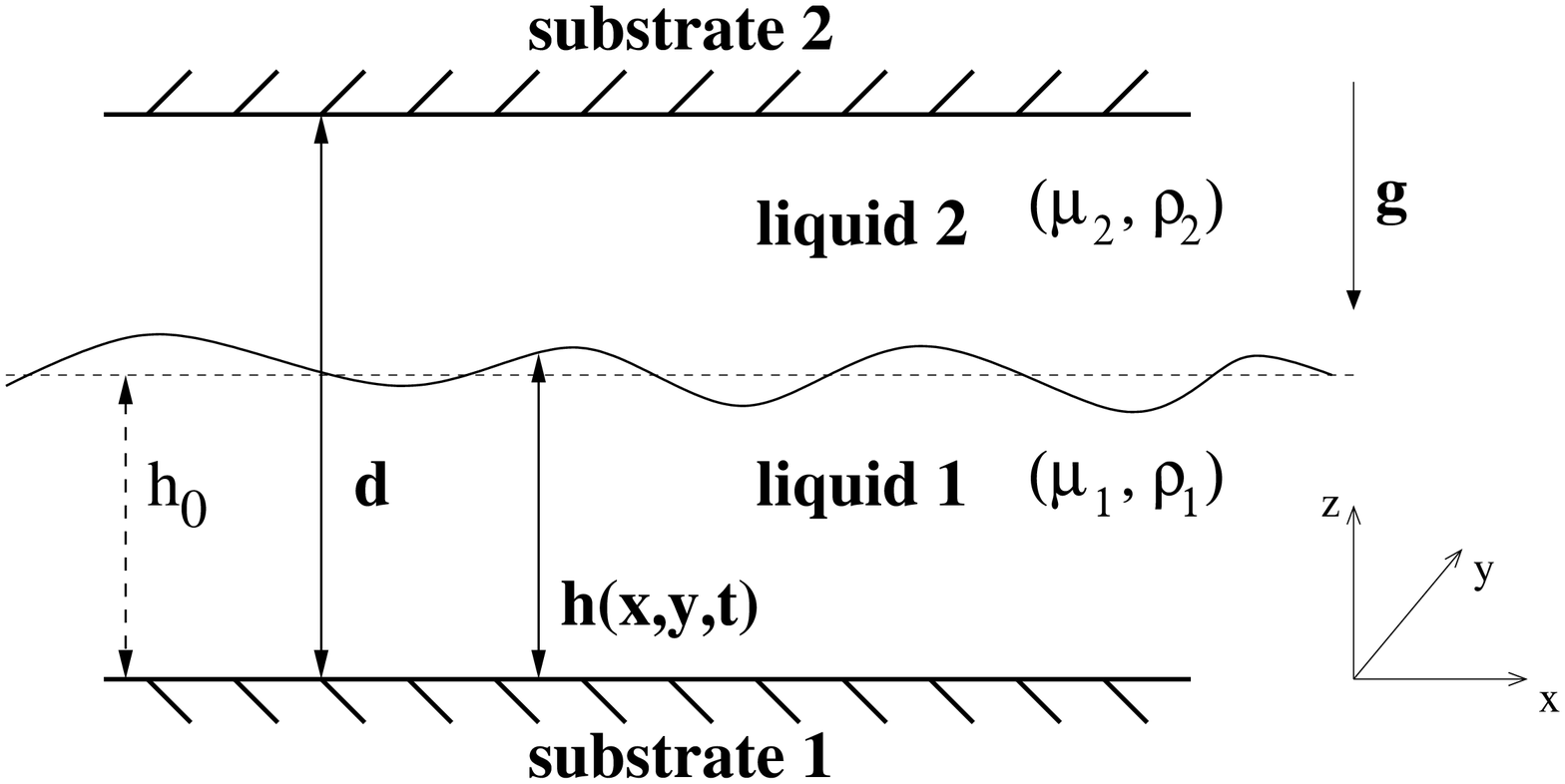}
\vfill
\caption{D.Merkt, Physics of Fluids}\label{fig1}
\end{center}
\end{figure}

\clearpage
\begin{figure}[ht]
\begin{center}
\includegraphics[width=\textwidth]{fig2.eps}
\vfill
\caption{D.Merkt, Physics of Fluids}\label{fig2} 
\end{center}
\end{figure}

\clearpage
\begin{figure}[ht]
\begin{center}
\includegraphics[width=\textwidth]{fig3.eps}
\vfill
\caption{D.Merkt, Physics of Fluids}\label{fig3} 
\end{center}
\end{figure}

\clearpage
\begin{figure}[ht]
\begin{center}
\includegraphics[width=\textwidth]{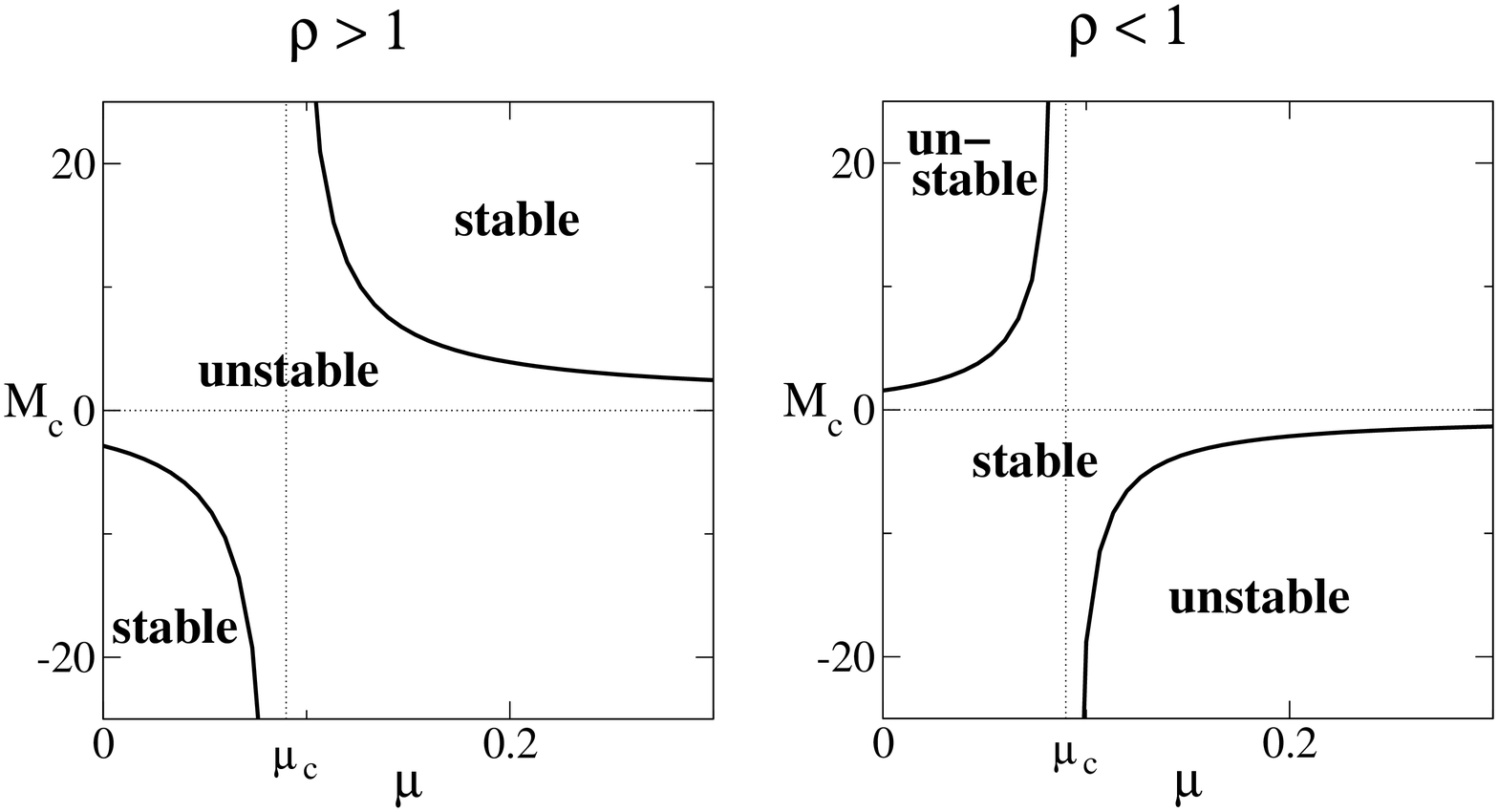}
\vfill
\caption{D.Merkt, Physics of Fluids}\label{fig4} 
\end{center}
\end{figure}

\clearpage
\begin{figure}[ht]
\begin{center}
\includegraphics[width=\textwidth]{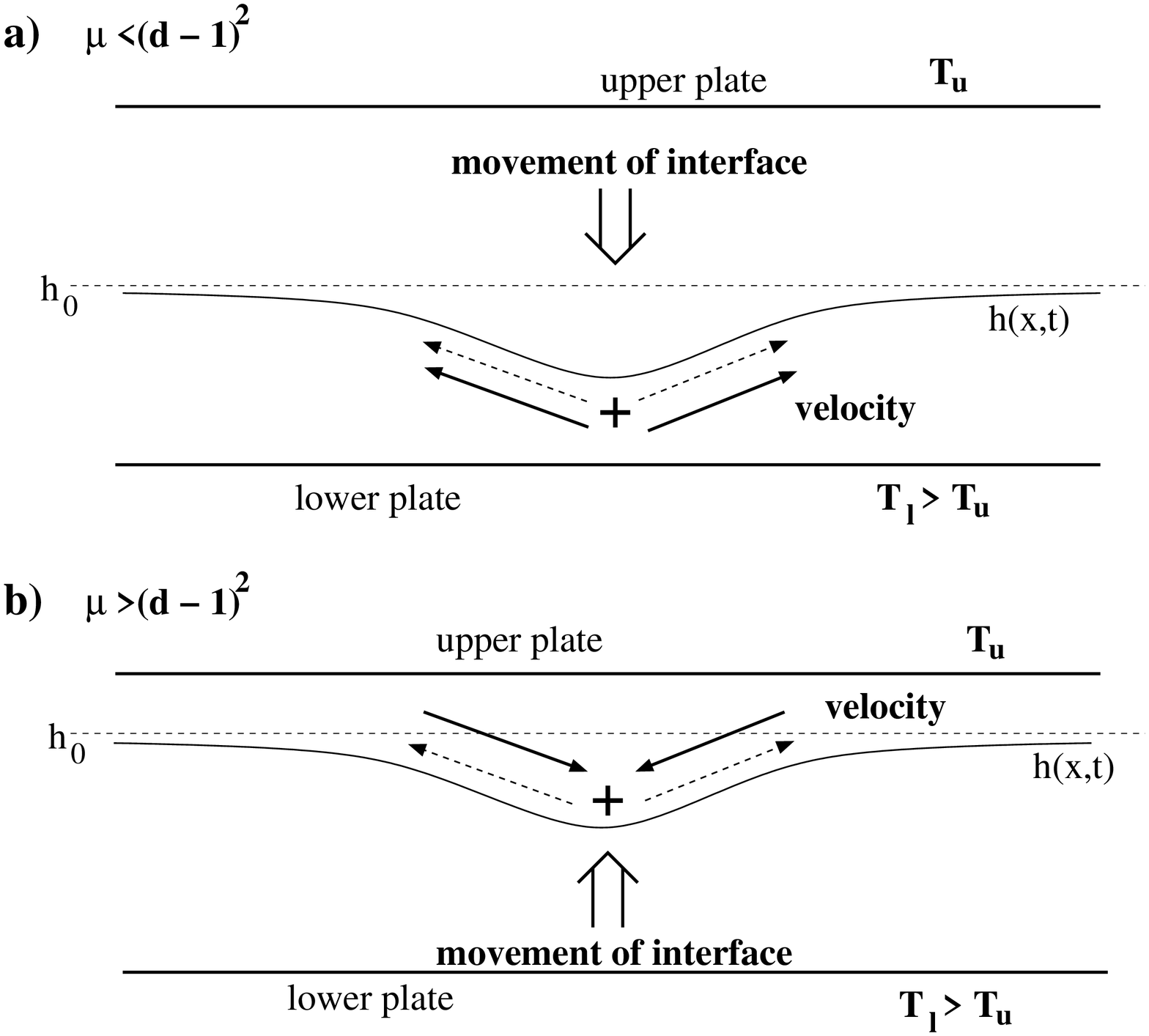}
\vfill
\caption{D.Merkt, Physics of Fluids}\label{fig5} 
\end{center}
\end{figure}

\clearpage
\begin{figure}[ht]
\begin{center}
\includegraphics[width=\textwidth]{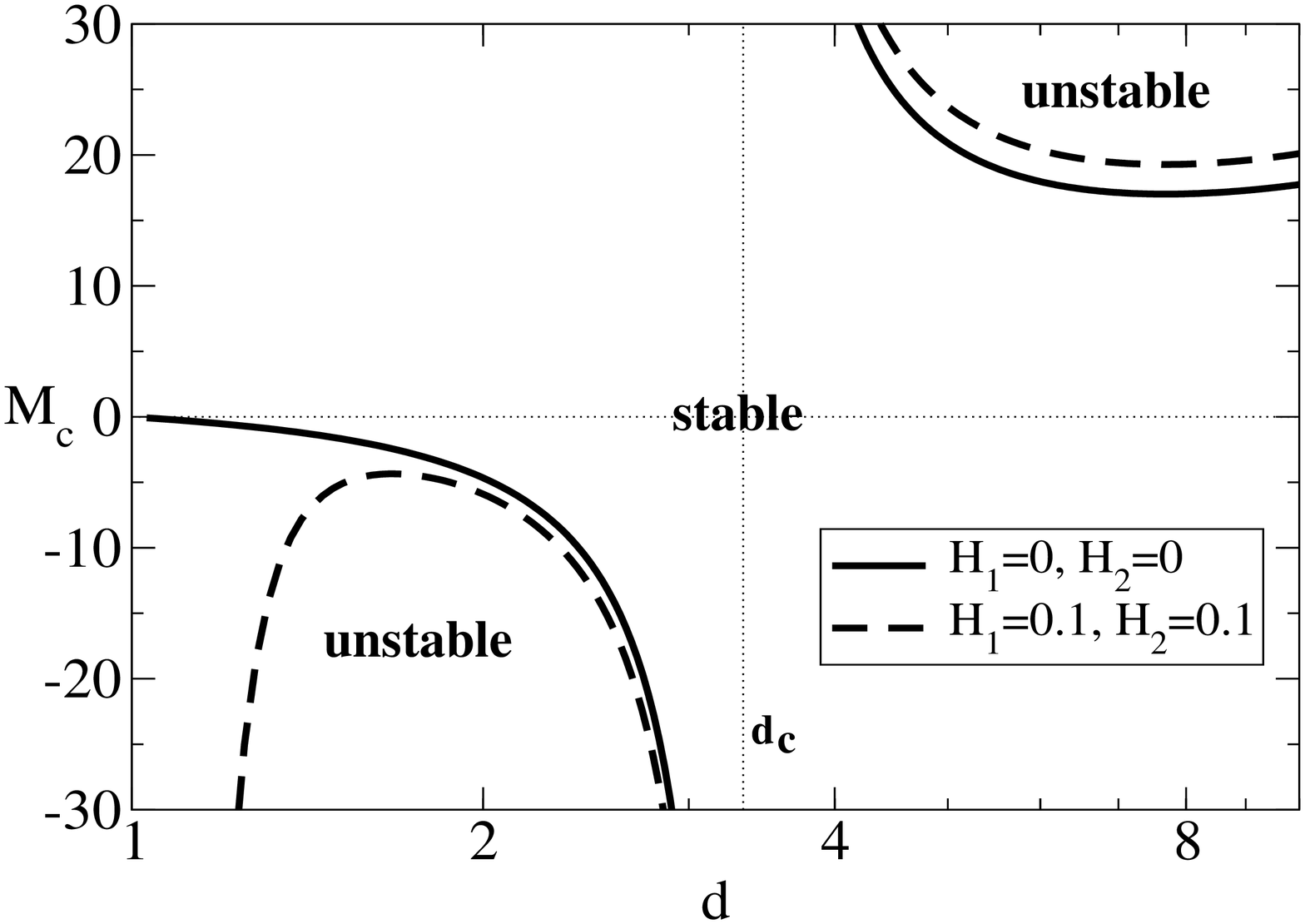}
\vfill
\caption{D.Merkt, Physics of Fluids}\label{fig6} 
\end{center}
\end{figure}

\clearpage
\begin{figure}[ht]
\begin{center}
\includegraphics[width=0.8\textwidth]{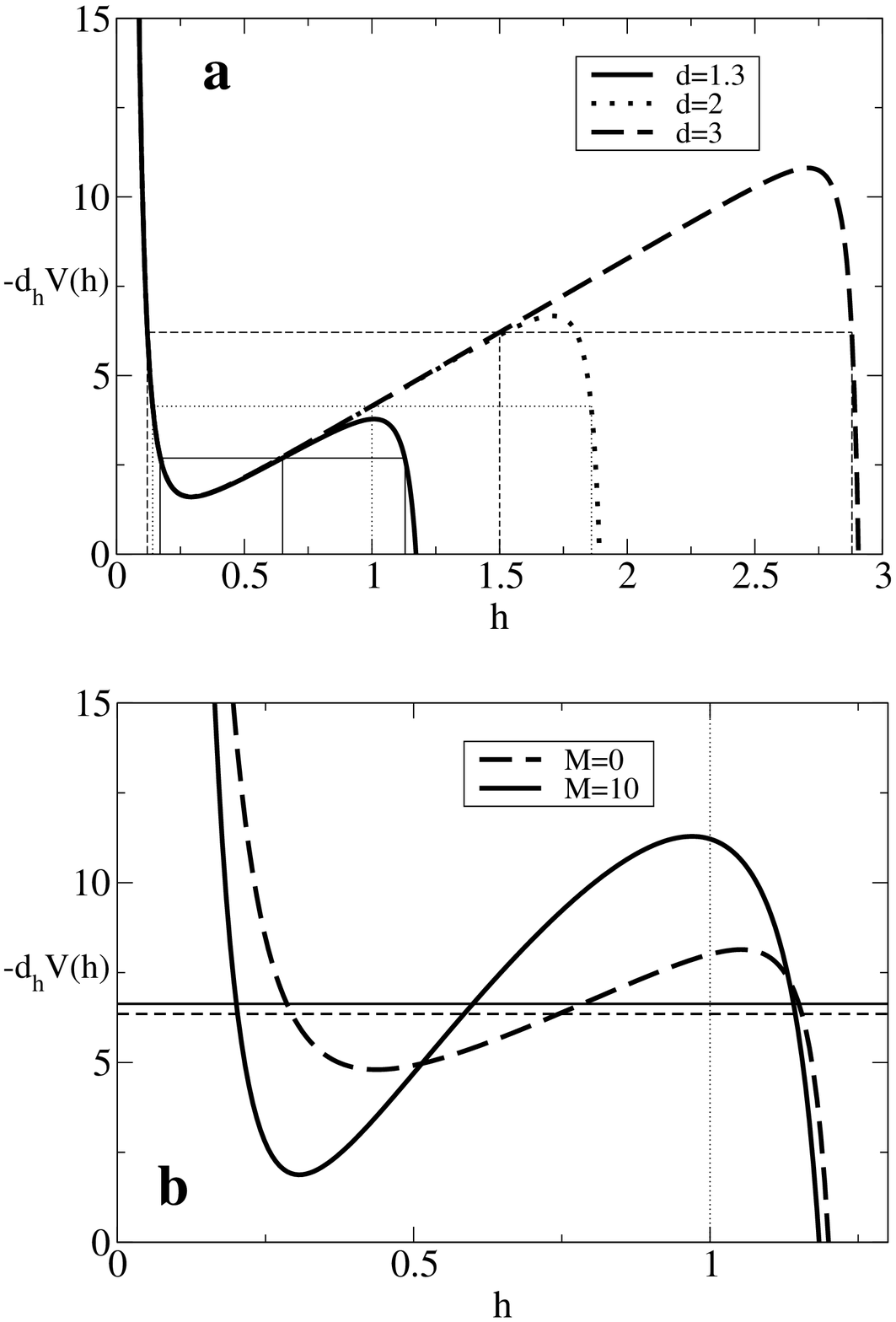}
\vfill
\caption{D.Merkt, Physics of Fluids}\label{fig7} 
\end{center}
\end{figure}

\clearpage
\begin{figure}[ht]
\begin{center}
\includegraphics[width=\textwidth]{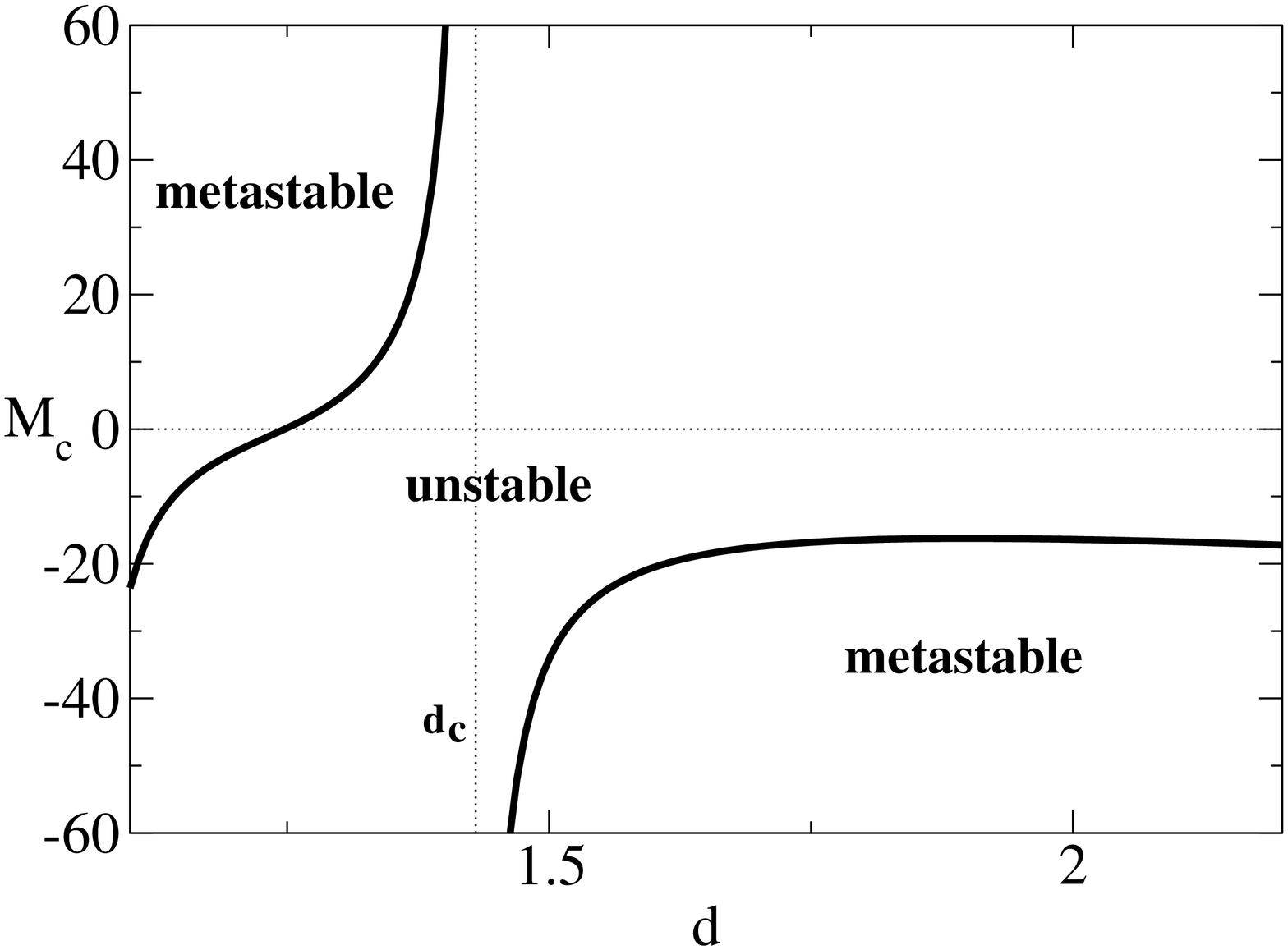}
\vfill
\caption{D.Merkt, Physics of Fluids}\label{fig_metas} 
\end{center}
\end{figure}

\clearpage
\begin{figure}[ht]
\begin{center}
\includegraphics[width=\textwidth]{fig9.eps}
\vfill
\caption{D.Merkt, Physics of Fluids}\label{figrt_m}
\end{center}
\end{figure}

\clearpage
\begin{figure}[ht]
\begin{center}
\includegraphics[width=\textwidth]{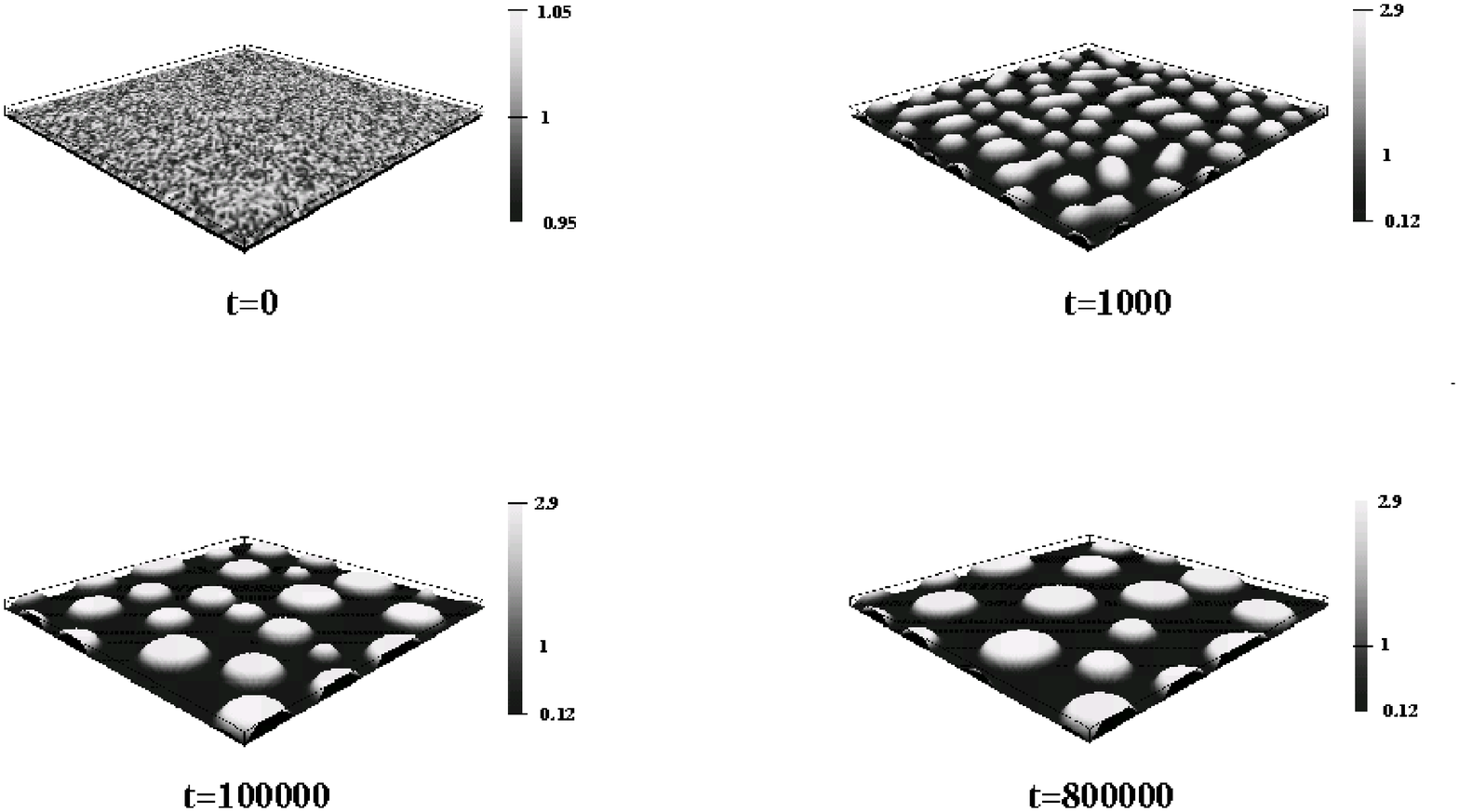}
\vfill
\caption{D.Merkt, Physics of Fluids}\label{fig_RT} 
\end{center}
\end{figure}

\clearpage
\begin{figure}[ht]
\begin{center}
\includegraphics[width=0.8\textwidth]{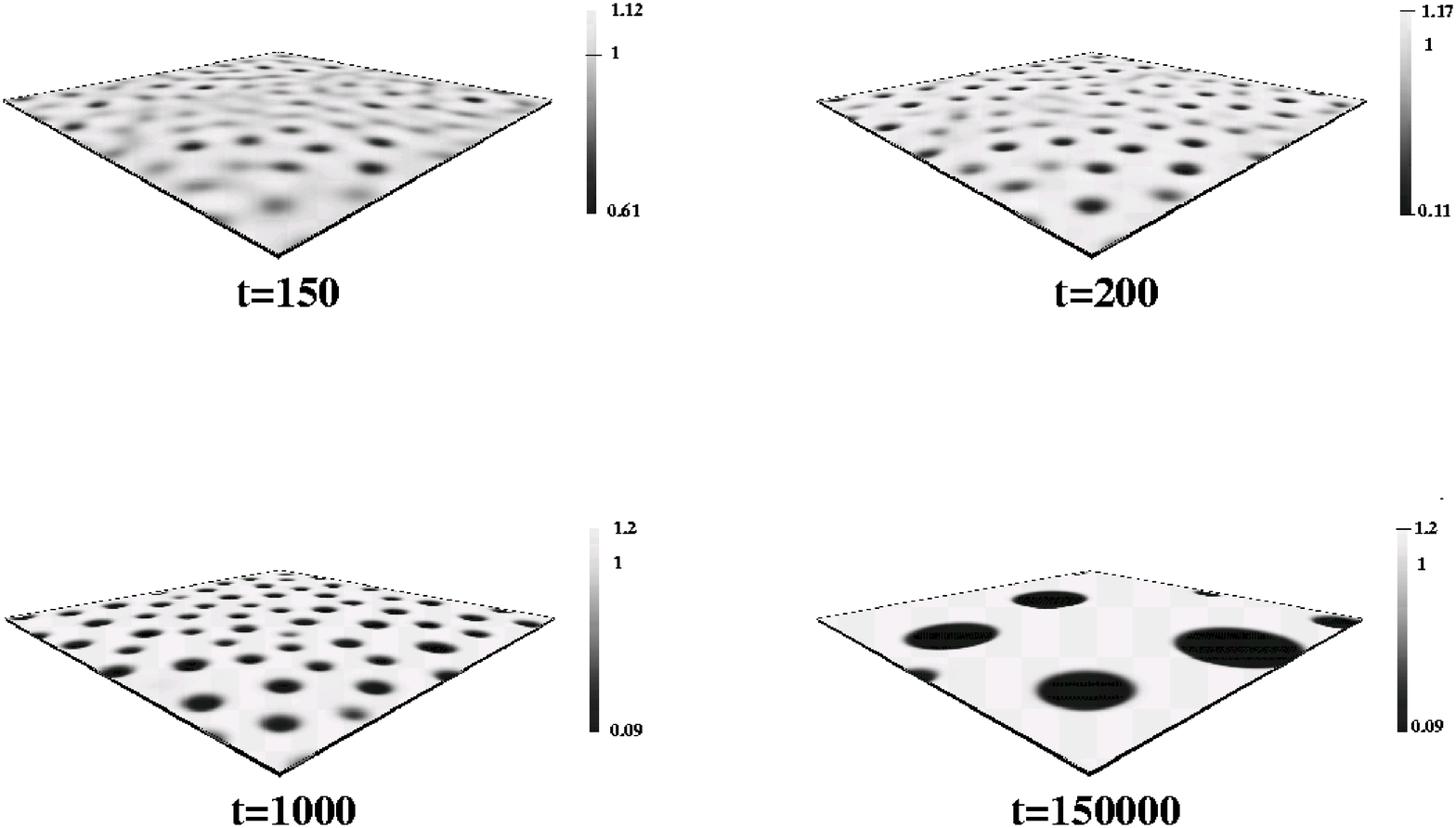}
\vfill
\caption{D.Merkt, Physics of Fluids}\label{fig_RT_2} 
\end{center}
\end{figure}

\clearpage
\begin{figure}[ht]
\begin{center}
\includegraphics[width=\textwidth]{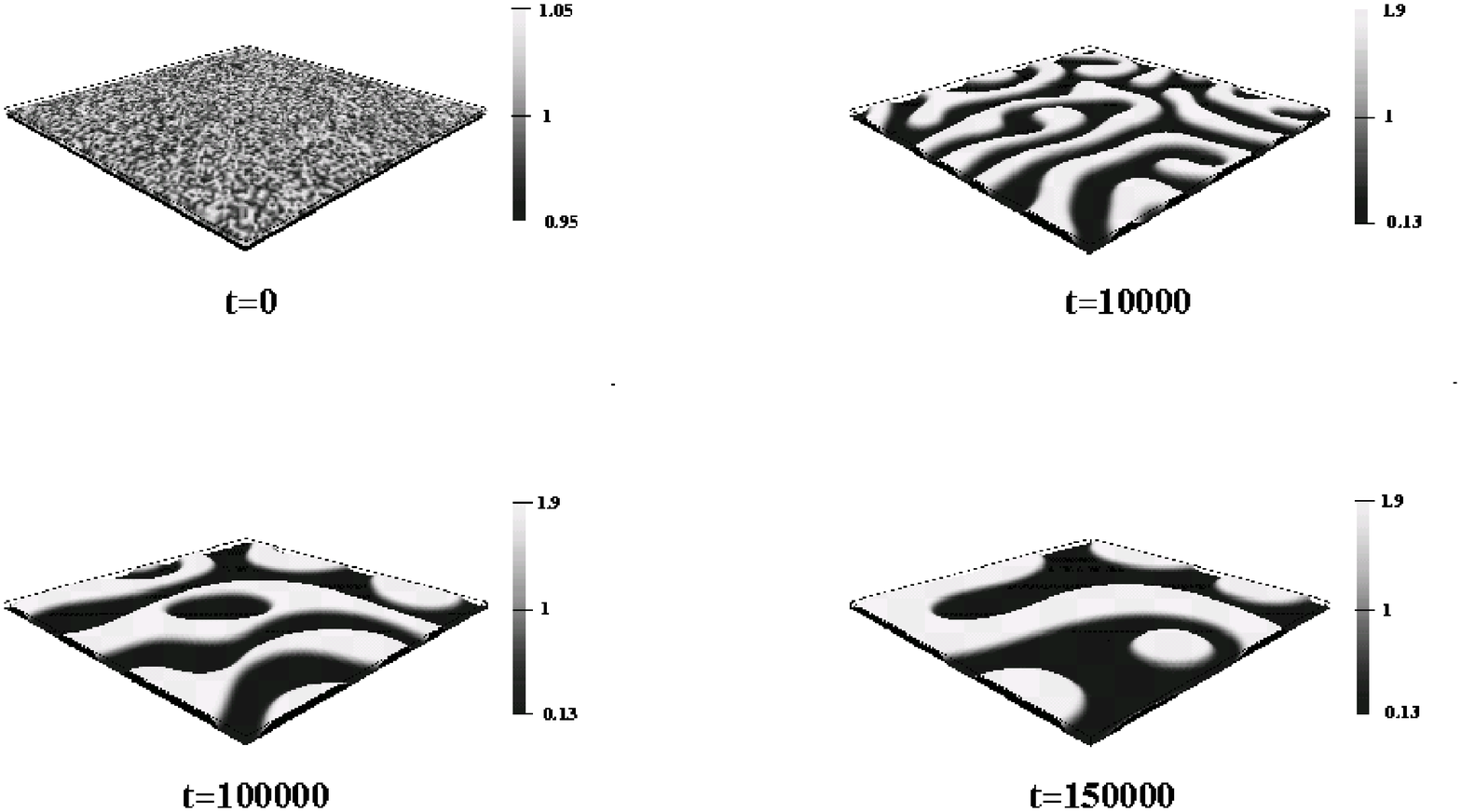}
\vfill
\caption{D.Merkt, Physics of Fluids}\label{fig_RT_M} 
\end{center}
\end{figure}

\clearpage
\begin{figure}[ht]
\begin{center}
\includegraphics[width=\textwidth]{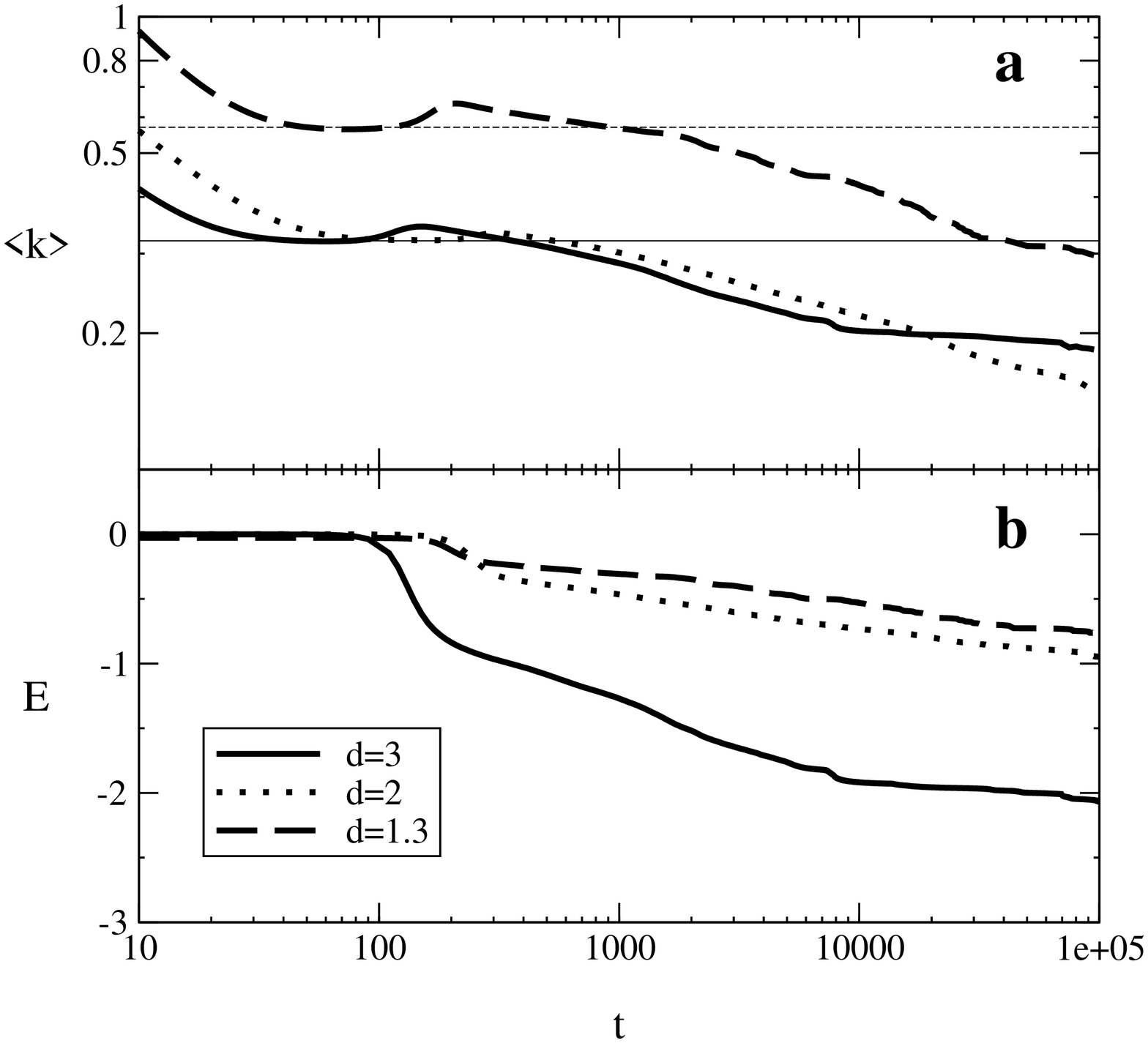}
\vfill
\caption{D.Merkt, Physics of Fluids}\label{fig_RT_K} 
\end{center}
\end{figure}

\clearpage
\begin{figure}[ht]
\begin{center}
\includegraphics[width=\textwidth]{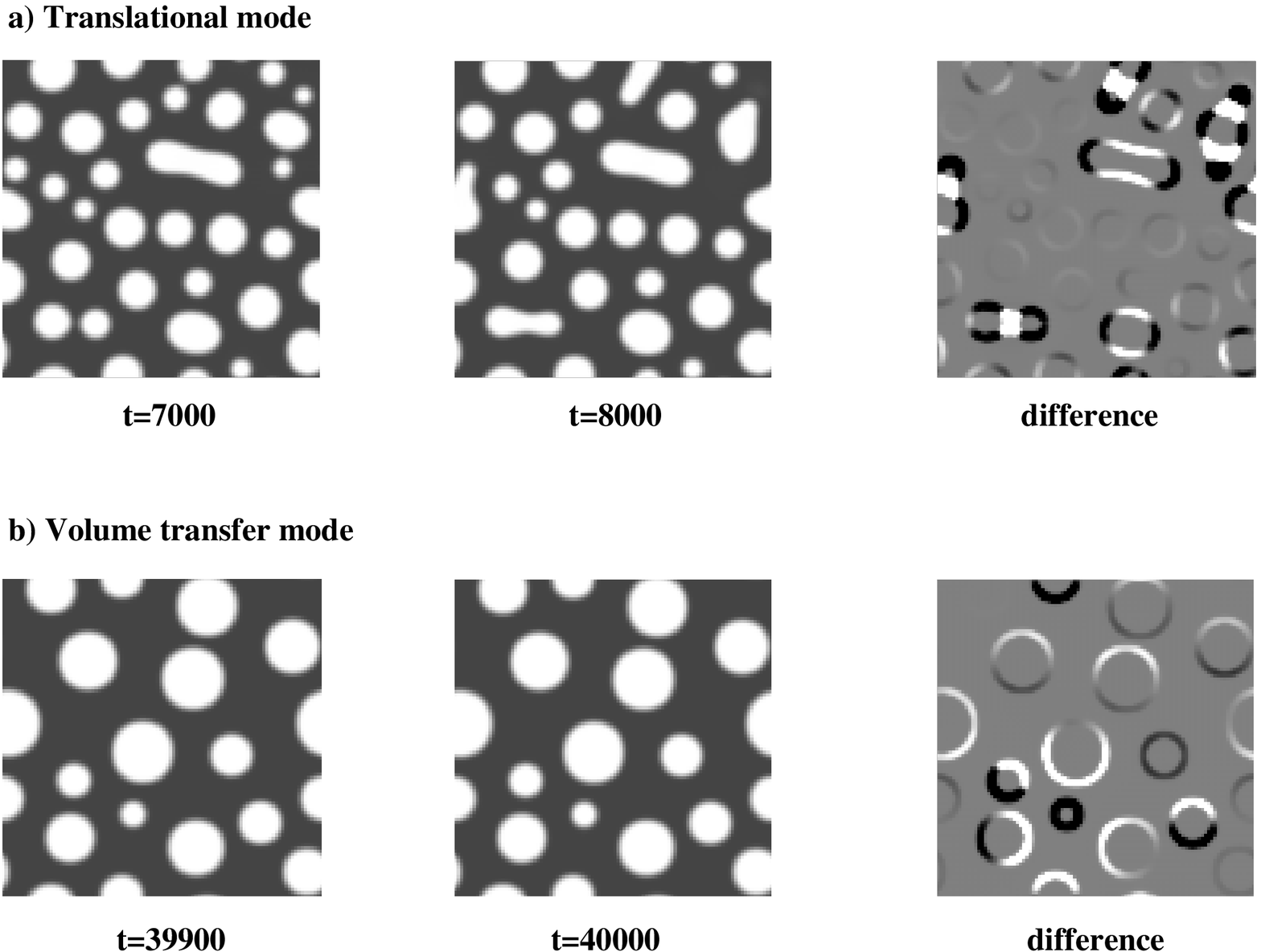}
\vfill
\caption{D.Merkt, Physics of Fluids}\label{fig_RT_diff} 
\end{center}
\end{figure}

\clearpage
\begin{figure}[ht]
\begin{center}
\includegraphics[width=\textwidth]{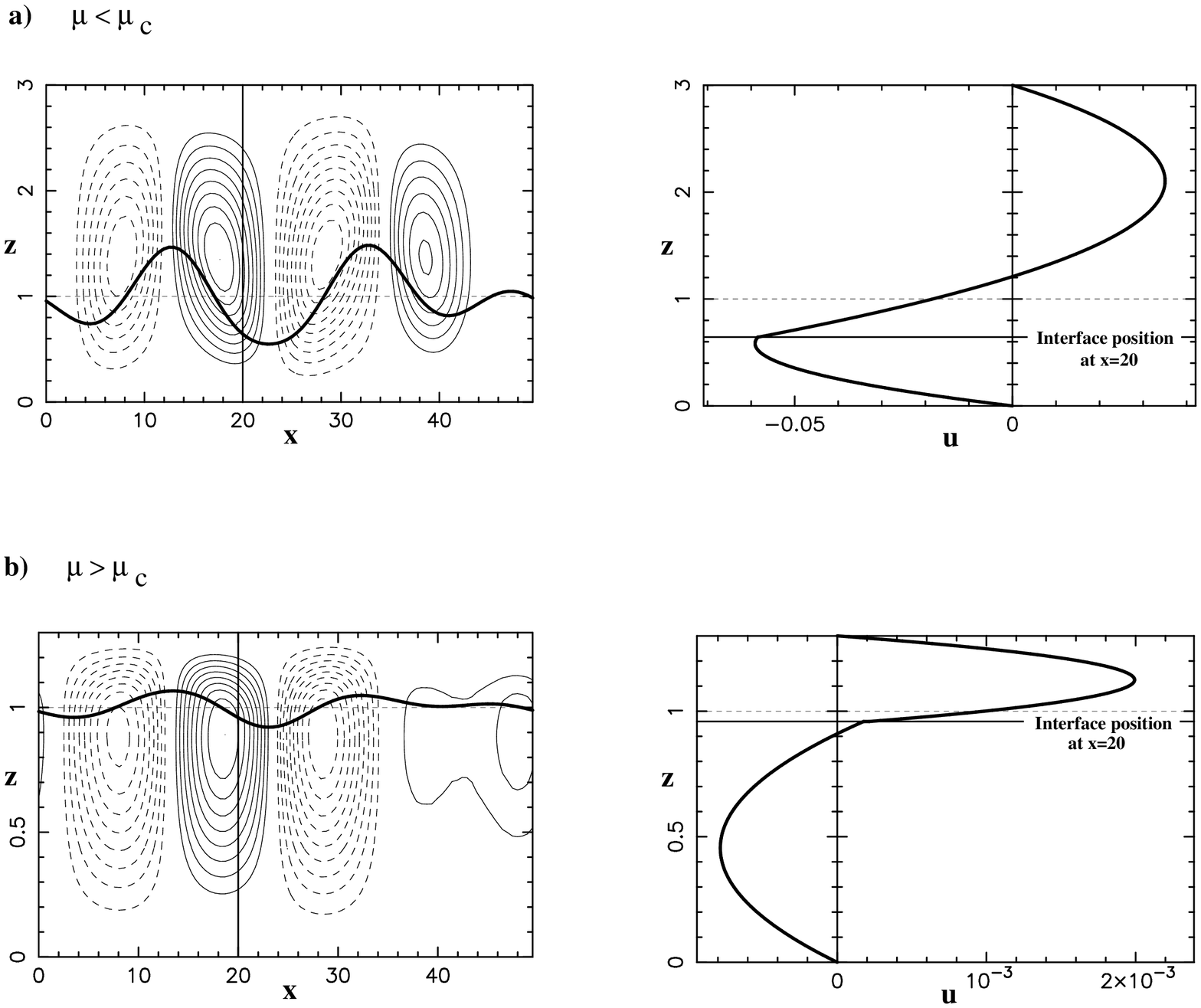}
\vfill
\caption{D.Merkt, Physics of Fluids}\label{figrt_vel} 
\end{center}
\end{figure}

\clearpage
\begin{figure}[ht]
\begin{center}
\includegraphics[width=\textwidth]{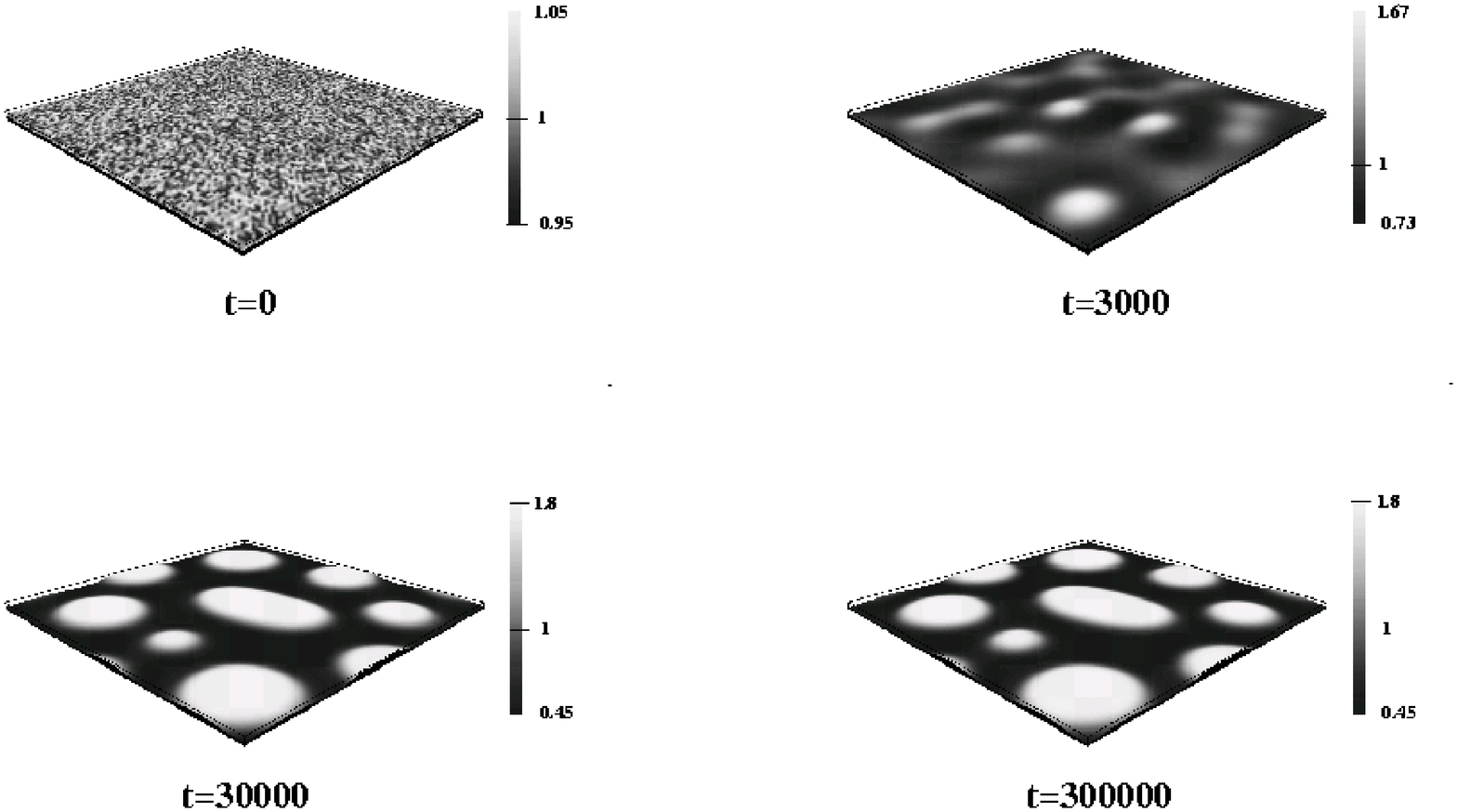}
\vfill
\caption{D.Merkt, Physics of Fluids}\label{fig_M_1} 
\end{center}
\end{figure}

\clearpage
\begin{figure}[ht]
\begin{center}
\includegraphics[width=\textwidth]{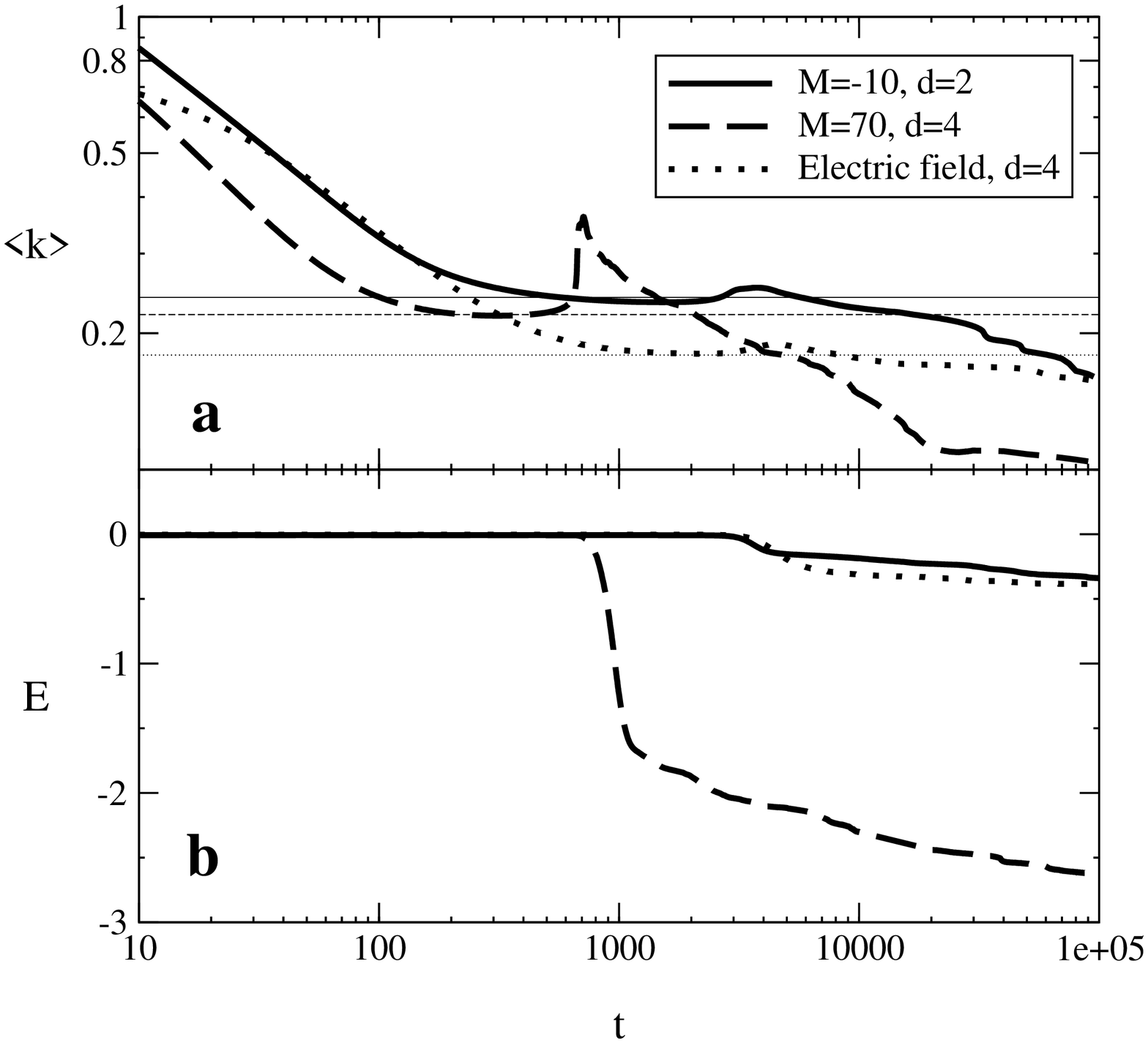}
\vfill
\caption{D.Merkt, Physics of Fluids}\label{fig_RT_K_1}
\end{center}
\end{figure}

\clearpage
\begin{figure}[ht]
\begin{center}
\includegraphics[width=\textwidth]{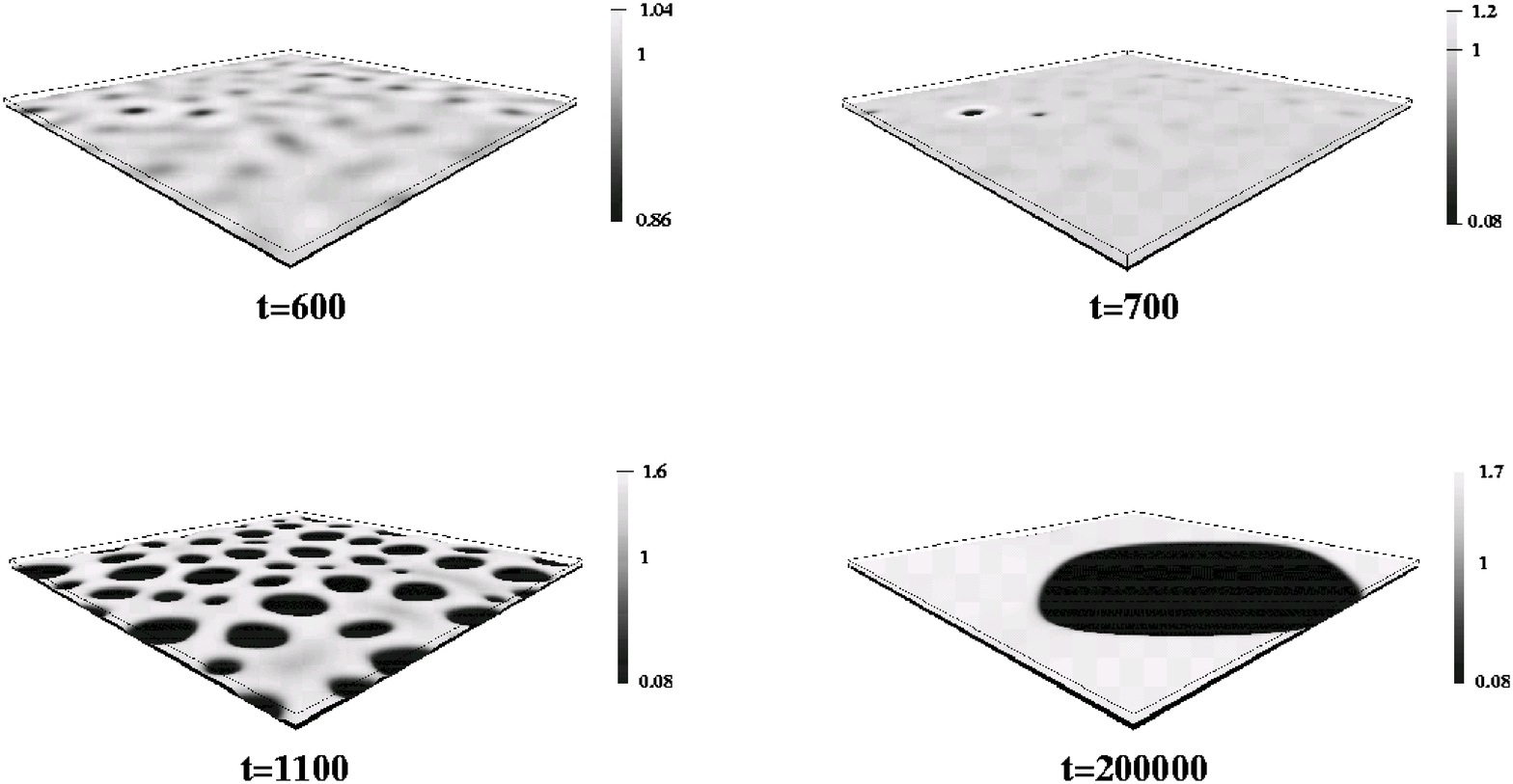}
\vfill
\caption{D.Merkt, Physics of Fluids}\label{fig_M} 
\end{center}
\end{figure}

\clearpage
\begin{figure}[ht]
\begin{center}
\includegraphics[width=\textwidth]{fig19.eps}
\vfill
\caption{D.Merkt, Physics of Fluids}\label{fig_MOB}
\end{center}
\end{figure}

\clearpage
\begin{figure}[ht]
\begin{center}
\includegraphics[width=\textwidth]{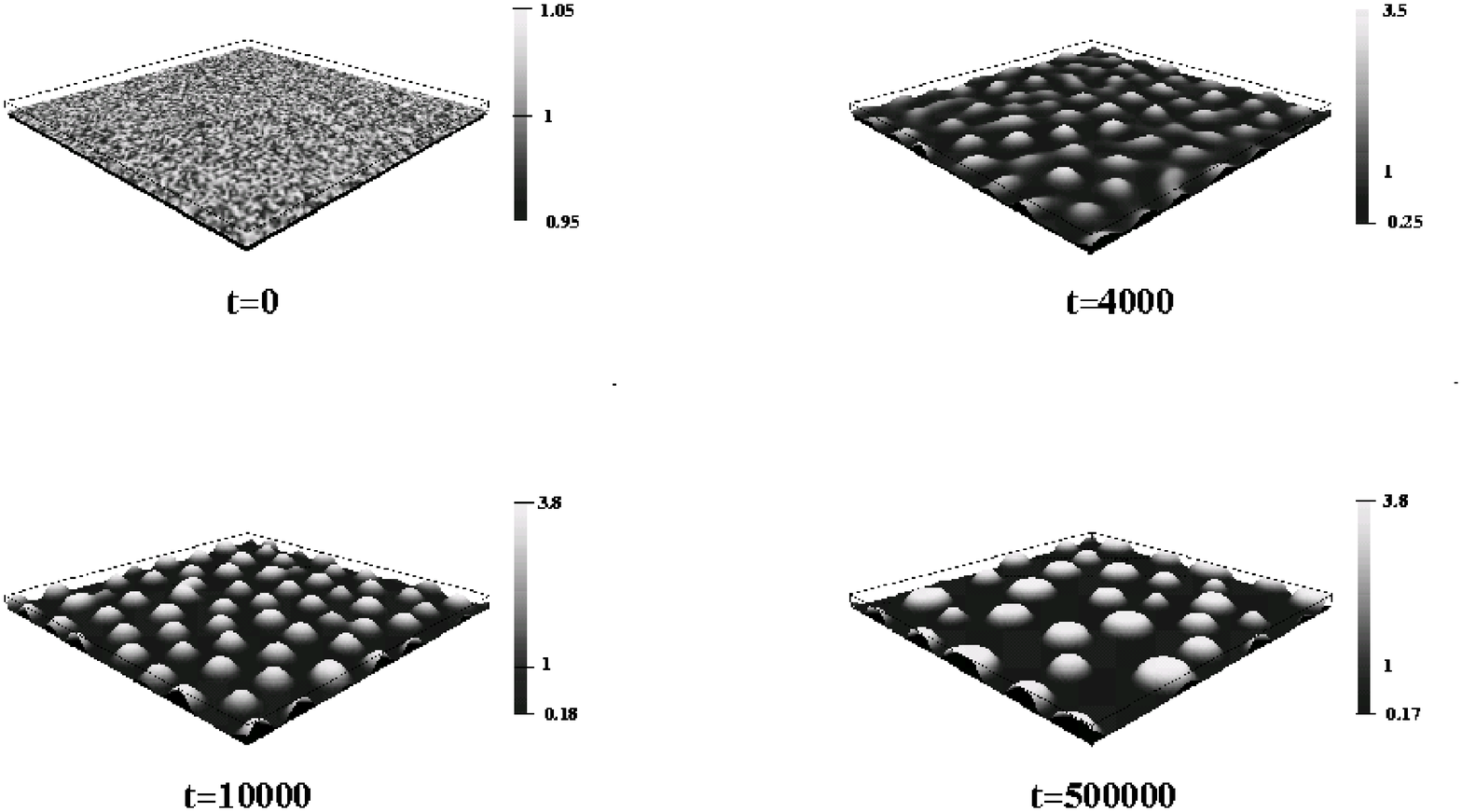}
\vfill
\caption{D.Merkt, Physics of Fluids}\label{fig_UC}
\end{center}
\end{figure}

\end{document}